\documentclass{article}

\usepackage{zharticle}
\usepackage{graphicx,amsmath,amsfonts,verbatim}
\usepackage{enumitem}
\usepackage{booktabs}
\usepackage{makecell}
\usepackage{multirow}
\usepackage[labelfont={bf,sf}, font=footnotesize]{subcaption}
\numberwithin{equation}{section}
\allowdisplaybreaks%
\usepackage{threeparttable}
\usepackage{rotating} 

\newcommand{\ones}{{\bf 1{}}}  
\newcommand{\reals}{{\bf R{}}}  
\newcommand{\ints}{{\bf Z{}}}  
\newcommand{\prob}{\mathop{\bf prob{}}}  
\newcommand{\expect}{\mathop{\bf E{}}}  
\newcommand{\diag}{\mathop{\bf diag}}  
\newcommand{\norm}[1]{{\lVert #1 \rVert}}  
\newcommand{\uni}{{\cal U{}}}  

\newcommand{\eg}{{\rmfamily\itshape e.g.}}
\newcommand{\ie}{{\rmfamily\itshape i.e.}}

\newcommand{\etal}{{\rmfamily\itshape et al.}}

\usepackage{lipsum}

\usepackage[colorlinks]{hyperref}

\usepackage{listings}
\usepackage{xcolor}

\definecolor{codegreen}{rgb}{0,0.6,0}
\definecolor{codegray}{rgb}{0.5,0.5,0.5}
\definecolor{codepurple}{rgb}{0.58,0,0.82}
\definecolor{backcolour}{rgb}{0.95,0.95,0.98}

\lstdefinestyle{zhstyle}{
    backgroundcolor=\color{backcolour},
    commentstyle=\color{codegreen},
    keywordstyle=\color{magenta},
    numberstyle=\tiny\color{codegray},
    stringstyle=\color{codepurple},
    basicstyle=\ttfamily\small,
    breakatwhitespace=false,
    breaklines=true,
    captionpos=b,
    keepspaces=true,
    numbers=left,
    numbersep=5pt,
    showspaces=false,
    showstringspaces=false,
    showtabs=false,
    tabsize=2,
}
\lstdefinelanguage{zhpython}{%
    sensitive=true,%
    morekeywords={and, as, assert, break, class, continue, def, del, elif,%
        else, except, exec, finally, for, from, global, if, import, in, is,%
        lambda, not, or, pass, print, raise, return, try, while, with, yield},%
    morecomment=[l]\#,%
    morestring=[s]{'''}{'''},%
    morestring=[s]{"""}{"""},%
    morestring=[b]',%
    morestring=[b]"%
}

\title{Fitting Reinforcement Learning Model to\\Behavioral Data under Bandits}
\author[1,2]{Hao Zhu}
\author[1,2]{Jasper Hoffmann}
\author[1,2]{Baohe Zhang}
\author[1,2]{Joschka Boedecker}
\affil[1]{IMBIT//BrainLinks-BrainTools}
\affil[2]{Department of Computer Science, University of Freiburg}

\begin{document}
\maketitle

\begin{abstract}
    We consider the problem of fitting a reinforcement learning (RL) model to some given behavioral data under a multi-armed bandit environment.
    These models have received much attention in recent years for characterizing human and animal decision making behavior.
    We provide a generic mathematical optimization problem formulation for the fitting problem of a wide range of RL models that appear frequently in scientific research applications.
    We then provide a detailed theoretical analysis of its convexity properties.
    Based on the theoretical results, we introduce a novel solution method for the fitting problem of RL models based on convex relaxation and optimization.
    Our method is then evaluated in several simulated and real-world bandit environments to compare with some benchmark methods that appear in the literature.
    Numerical results indicate that our method achieves comparable performance to the state-of-the-art, while significantly reducing computation time.
    We also provide an open-source Python package for our proposed method to empower researchers to apply it in the analysis of their datasets directly, without prior knowledge of convex optimization.
\end{abstract}

\vspace*{\fill}
\begin{center}\sffamily
    This manuscript has been accepted for publication in \emph{Frontiers in Applied Mathematics and Statistics}.
    This preprint version should include all the content (or possibly a slightly more) in the final published version, but may differ in formatting and copyediting.
\end{center}
\vspace*{\fill}

\newpage
\tableofcontents
\newpage

\section{Introduction}
We consider the problem of fitting a reinforcement learning (RL) model to some given behavioral data under a multi-armed bandit environment.

\paragraph{Bandits and reinforcement learning.}
A \emph{bandit problem} is a sequential game between a player and an environment, where the player may choose one of several actions at each time step, and receives some reward from the environment that depends on the selected action.
The actions are often referred to as \emph{arms} in the literature, so a $k$-armed bandit refers to a bandit problem with $k$ possible actions~\cite{lattimore2020bandit}.
The player's goal is to maximize the cumulative reward across the whole episode, which requires the player to learn the reward structure of the environment and make informed decisions based on the learned information.
Obviously, the player under the bandit setting can only select their action based on the history of their past actions and the received rewards, which forms a typical setting for RL problems~\cite{sutton2018reinforcement}.
Multi-armed bandits has been a very active research area of engineering, including computer science, operations research, economics, statistics, etc.~\cite{slivkins2019introduction}.

\paragraph{Multi-armed bandit behavioral tasks.}
Apart from the applications in engineering, the class of multi-armed bandit tasks is also an experimental paradigm used to investigate a wide range of animals' decision making processes.
These tasks have been widely used in, \eg, neuroscience~\cite{samejima2005representation,tai2012transient,parker2016reward,donahue2018distinct,ebitz2018exploration,miller2018predictive,bari2019stable,costa2019subcortical,hattori2019area,hamaguchi2022prospective,de2023freibox,hattori2023meta,zhu2024multiintention}, psychology~\cite{daw2006cortical,vertechi2020inference,kleespies2025sleep}, and medical~\cite{seymour2016deep} researches.
In these tasks, the animal or human subject is faced with multiple choices with different rewards, and may choose one of them for each trial.
Some variants also include shuffling the reward assigned to each choice regularly or randomly (which are sometimes referred to as \emph{dynamic bandits}), or introducing an external cue, \eg, image, sound, etc., to individual choices.
Under the latter type of environments, subjects can select their action according to some contextual information, instead of just based on trial and error.
Nevertheless, the common goal of the bandit task for the subject is to maximize the cumulative reward across the whole episode (\ie, experimental session).

\paragraph{RL model for decision making under bandits.}
In the context of engineering, RL is a class of algorithms that can be used for solving multi-armed bandit problems.
Meanwhile, in behavioral science, RL also serves as one of the mathematical models for characterizing the decision making behavior of animals and human under a bandit task.
(See, \eg, Beron~\etal~\cite{beron2022mice} for a review of different models for animal behavior characterization.)
The following procedure describes the most basic instance of the class of RL models, namely the \emph{forgetting Q-learning} model~\cite{ito2009validation,beron2022mice}:
Let $m \in \ints_{++}$ be the number of possible \emph{actions} (choices) in the bandit task, and let $t \in \ints_+$ be the discrete time step.
After the choice at time step $t-1$, the subject receives a \emph{reward signal} $u(t) \in \reals^m$ that depends on the selected action, given by
\begin{equation}\label{eq:fql_rewsig}
    u_i(t) = \left\{
        \begin{array}{ll}
            1 & \mbox{if action $i$ was selected \textit{and} rewarded}\\
            0 & \mbox{otherwise}
        \end{array}\right.
\end{equation}
for $i = 1, \ldots, m$.
To maximize the cumulative reward, the subject formulates some \emph{value function} (or really, vector) $x(t) \in \reals^m$ for each time step $t \geq 1$, and recursively updates it according to
\begin{equation}\label{eq:fql_v}
    x(t) = x(t - 1) + \alpha(\beta u(t) - x(t - 1)),
\end{equation}
where the parameters $\beta \in [0, \infty)$ can be interpreted as the \emph{sensitivity} to the reward signal $u(t)$, and $\alpha \in [0, 1]$ is the \emph{learning rate} of the value estimation error $(\beta u(t) - x(t - 1))$.
By convention, the initial value function at $t = 0$ is set to $x(0) = 0$.
Let $a(t) \in \{1, \ldots, m\}$ denote the subject's action at the $t$th time step, which is then assumed to be selected according to:
\begin{equation}\label{eq:fql_pi}
    \prob(a(t) = i) = \frac{\exp(x_i(t))}{\sum_{j = 1}^{m} \exp(x_j(t))},\quad i = 1, \ldots, m,\quad t = 0, \ldots, n.
\end{equation}
In addition to the basic example above, several extensions to the forgetting Q-learning model have been developed to capture more subtle behavioral properties in modeling (see \S\ref{sec:fql_fit_ext} for more details).

\paragraph{RL model fitting problem.}
Behavioral science researchers, \ie, the users of the RL models, are interested in obtaining individual subject's behavior characterization, given the observed behavior outcome.
In particular, for the case of the forgetting Q-learning model defined by (\ref{eq:fql_rewsig}) to (\ref{eq:fql_pi}), the goal is to recover the model parameters $\alpha$ and $\beta$ as well as the value functions $x(t)$, $t = 1, \ldots, n$, given the dataset ${\{(u(t), a(t))\}}_{t = 1}^n$ (although in practice the value functions are generally of more interest).
Roughly speaking, this leads to solving an optimization problem with the objective function being the likelihood of observing the subject's behavior under the model assumptions.
The variables for these problems are then the model parameters and the value function at each time step.
(A formal definition of the RL model fitting problem will be introduced in \S\ref{sec:fit_prob_rlm}.)

\paragraph{This paper.}
Despite the fast growing of RL behavior model applications in the scientific research community, a generic formal definition and analysis of the properties for the RL model fitting problem have not yet been well established.
As a partial consequence, regarding the practical aspect, current solution methods for fitting RL models are either very slow or difficult to implement and debug (see \S\ref{sec:expt} for more detail).
In this paper, we aim at filling these gaps with the following three folds of contributions:
\begin{itemize}
    \item Firstly, we formalize the mathematical optimization problem corresponding to the fitting problem of a wide range of the most commonly used RL models in \S\ref{sec:fit_prob_rlm}.
    \item Then, in \S\ref{sec:cvx_alys}, we evaluate the convexity properties of the RL model fitting problems according to our problem formulation.
    \item Finally, based on the theoretical results, in \S\ref{sec:solution_cvx}, we introduce a novel solution method for fitting RL models to bandit behavioral data via convex optimization.
\end{itemize}
Our proposed solution method is then evaluated in several simulated and real-world bandit environments to compare with some benchmark methods that appear in the literature.
Numerical results indicate that our method achieves comparable performance with significantly decreased computing time.
The implementation of our method is fully open-sourced as a Python package under
\begin{quote}
    \url{https://github.com/nrgrp/rlfit},
\end{quote}
such that it can be easily applied by users not well versed in convex analysis and optimization.

\section{The fitting problem of RL models}\label{sec:fit_prob_rlm}
\subsection{Basic forgetting Q-learning model}\label{sec:forget_qlearning}
We start from the fitting problem of the basic forgetting Q-learning model given by (\ref{eq:fql_rewsig}) to (\ref{eq:fql_pi}).
Recall that here the objective is to maximize the likelihood of observing the given data $a(t)$ and $u(t)$ for $t = 1, \ldots, n$, with the variables being the value functions $x(t)$ and the model parameters $\alpha$ and $\beta$.
First, notice that (\ref{eq:fql_v}) can be written as
\begin{equation*}
    x(t) = (1 - \alpha)x(t-1) + \alpha\beta u(t).
\end{equation*}
For simplicity of notation, we transform the actions $a(t) \in \{1, \ldots, m\}$ into one-hot representations, given by $y(t) \in \{e_1, \ldots, e_m\} \subseteq \reals^m$, where $e_i$ is the $i$th standard basis vector, \ie,
\begin{equation}\label{eq:y_t}
    y_i(t) = \left\{
        \begin{array}{ll}
            1 & a(t) = i\\
            0 & \mbox{otherwise}
        \end{array}\right.
\end{equation}
for all $i = 1, \ldots, m$.
Then the log-likelihood of observing $a(t)$ at time step $t$ is
\begin{equation}\label{eq:ll}
    \ell(x(t), y(t)) = \log \left({y(t)}^T \left(\frac{\exp(x(t))}{\sum_{i=1}^m \exp(x_i(t))}\right)\right).
\end{equation}
Put together, the fitting problem of the forgetting Q-learning model can be written as
\begin{equation}\label{prob:fql}
    \begin{array}{ll}
        \mbox{minimize} & -\sum_{t=1}^{n} \ell(x(t), y(t))\\
        \mbox{subject to} & x(t) = (1 - \alpha) x(t - 1) + \alpha\beta u(t),\quad t = 1, \ldots, n\\
        & x(0) = 0,\quad 0 \leq \alpha \leq 1,\quad \beta \geq 0,
    \end{array}
\end{equation}
where the problem variables are $\alpha, \beta \in \reals$ and $x(1), \ldots, x(n) \in \reals^m$, the problem data are $y(t), u(t) \in \reals^m$, and each log-likelihood term in the objective is given by (\ref{eq:ll}).

\subsection{Extensions}\label{sec:fql_fit_ext}
\paragraph{Extension on reward signals.}
As a simple extension to the forgetting Q-learning model, one may assume a different reward signal $u(t)$ as in (\ref{eq:fql_rewsig}).
For example, some kind of `punishment' could possibly be integrated to the unrewarded choices, \eg, replacing all the zeros in (\ref{eq:fql_rewsig}) with $-1$~\cite{aylward2019altered,maboudi2020honeybees}.
This type of extensions only changes the problem data of (\ref{prob:fql}), which does not influence the properties of the fitting problem itself.

\paragraph{Multiple learning rates and reward sensitivity.}
One of the widely applied extensions to the basic forgetting Q-learning model is to incorporate different learning rates $\alpha$ and reward sensitivity $\beta$ for individual actions of the bandit~\cite{hattori2019area, hattori2023meta}.
In other words, for each entry $x_i(t)$ of the value function $x(t) \in \reals^m$, $i = 1, \ldots, m$, we have
\begin{equation*}
    x_i(t) = x_i(t - 1) + \alpha_i(\beta_i u_i(t) - x_i(t - 1)).
\end{equation*}
In this case, the model parameters $\alpha$ and $\beta$ are not just two real numbers, but two vectors in $\reals^m$ with each entry satisfying $\alpha_i \in [0, 1]$ and $\beta_i \in [0, \infty)$, $i = 1, \ldots, m$.
Hence, the model fitting problem (\ref{prob:fql}) is now extended to be
\begin{equation}\label{prob:multi_lr}
    \begin{array}{ll}
        \mbox{minimize} & -\sum_{t=1}^{n} \ell(x(t), y(t))\\
        \mbox{subject to} & x(t) = \diag(1 - \alpha) x(t - 1) + \diag(\alpha) \diag(\beta) u(t),\quad t = 1, \ldots, n\\
        & x(0) = 0,\quad 0 \preceq \alpha \preceq 1,\quad \beta \succeq 0
    \end{array}
\end{equation}
with variables $\alpha, \beta \in \reals^m$ and $x(1), \ldots, x(n) \in \reals^m$.

\paragraph{Subreward signals and subvalue functions.}
Another type of extension to the basic forgetting Q-learning model that appears frequently in practice takes the following assumption:
There exist multiple \emph{subreward signals} $u^{(1)}(t), \ldots, u^{(k)}(t) \in \reals^m$~\cite{beron2022mice}, corresponding to multiple \emph{subvalue functions} $z^{(1)}(t), \ldots, z^{(k)}(t) \in \reals^m$.
These subvalue functions are then updated individually with parameters $\alpha^{(i)}, \beta^{(i)} \in \reals$, according to
\begin{equation*}
    z^{(i)}(t) = z^{(i)}(t - 1) + \alpha^{(i)}(\beta^{(i)} u^{(i)}(t) - z^{(i)}(t - 1))
\end{equation*}
for all $i = 1, \ldots, k$.
The value function $x(t)$ used in (\ref{eq:fql_pi}) for action selection is then a linear combination of $z^{(1)}(t), \ldots, z^{(k)}(t)$ under some \textit{given} weight vector $w \in \reals^k$ (which is commonly assumed to be $w = \ones$), \ie, $x(t) = w_1 z^{(1)}(t) + \cdots + w_k z^{(k)}(t)$.
In this setup, the problem (\ref{prob:fql}) now becomes
\begin{equation}\label{prob:subrew}
    \begin{array}{ll}
        \mbox{minimize} & -\sum_{t=1}^{n} \ell(x(t), y(t))\\[5pt]
        \mbox{subject to} & x(t) = \left[
        \begin{array}{ccc}
            z^{(1)}(t) & \cdots & z^{(k)}(t)
        \end{array}\right] w\\[5pt]
        & z^{(i)}(t) = (1 - \alpha^{(i)}) z^{(i)}(t-1) + \alpha^{(i)} \beta^{(i)} u^{(i)}(t)\\
        & z^{(i)}(0) = 0,\quad 0 \leq \alpha^{(i)} \leq 1,\quad \beta^{(i)} \geq 0\\
        &i = 1, \ldots, k,\quad t = 1, \ldots, n,
    \end{array}
\end{equation}
where the variables are $\alpha^{(i)}, \beta^{(i)} \in \reals$, $z^{(i)}(1), \ldots, z^{(i)}(n) \in \reals^m$ for all $i = 1, \ldots, k$, and $x(1), \ldots, x(n) \in \reals^m$; the problem data are $w \in \reals^k$ and $y(t), u^{(i)}(t) \in \reals^m$, $t = 1, \ldots, n$, $i = 1, \ldots, k$.

\subsection{RL model fitting problems in general form}\label{sec:rlm_fit_gen}
It is easily seen that the RL model fitting problems, given by (\ref{prob:fql}), (\ref{prob:multi_lr}), and (\ref{prob:subrew}), can be written in the following general form:
\begin{equation}\label{prob:rl_fit}
    \begin{array}{ll}
        \mbox{minimize} & -\sum_{t=1}^{n} \ell(x(t), y(t))\\[5pt]
        \mbox{subject to} & x(t) = \left[
        \begin{array}{ccc}
            z^{(1)}(t) & \cdots & z^{(k)}(t)
        \end{array}\right] w\\[5pt]
        & z^{(i)}(t) = \diag(1 - \alpha^{(i)}) z^{(i)}(t-1) + \diag(\alpha^{(i)}) \diag(\beta^{(i)}) u^{(i)}(t)\\
        & z^{(i)}(0) = 0,\quad 0 \preceq \alpha^{(i)} \preceq 1,\quad \beta^{(i)} \succeq 0\\
        & i = 1, \ldots, k,\quad t = 1, \ldots, n,
    \end{array}
\end{equation}
where the variables are $\alpha^{(i)}, \beta^{(i)} \in \reals^m$, $z^{(i)}(1), \ldots, z^{(i)}(n) \in \reals^m$ for all $i = 1, \ldots, k$, and $x(1), \ldots, x(n) \in \reals^m$; the problem data are $w \in \reals^k$, $y(t), u^{(i)}(t) \in \reals^m$, $t = 1, \ldots, n$, $i = 1, \ldots, k$.
Assuming $w = \ones$, by taking $k = 1$, the problem (\ref{prob:rl_fit}) reduces to (\ref{prob:multi_lr});
by adding additional constraints $\alpha^{(i)}_1 = \cdots = \alpha^{(i)}_m$ and $\beta^{(i)}_1 = \cdots = \beta^{(i)}_m$, $i = 1, \ldots, k$, the problem (\ref{prob:rl_fit}) reduces to (\ref{prob:subrew});
and by combining the two additional requirements above together, we have the basic forgetting Q-learning model fitting problem (\ref{prob:fql}).
The time complexity of \emph{evaluating} the objective and constraints of (\ref{prob:rl_fit}) is $O(mnk)$.

\section{Convexity properties}\label{sec:cvx_alys}
To analyze the convexity properties of the general RL model fitting problem (\ref{prob:rl_fit}), we start by eliminating the recursive expression about $z^{(i)}(t)$.
Consider the $j$th entry of the vectors $z^{(i)}(0), \ldots, z^{(i)}(n)$, we have
\begin{align*}
    &z^{(i)}_j(0) = 0\\
    &z^{(i)}_j(1) = (1 - \alpha^{(i)}_j)z^{(i)}_j(0) + \alpha^{(i)}_j \beta^{(i)}_j u^{(i)}_j(1) = \alpha^{(i)}_j \beta^{(i)}_j u^{(i)}_j(1)\\
    &z^{(i)}_j(2) = (1 - \alpha^{(i)}_j)z^{(i)}_j(1) + \alpha^{(i)}_j \beta^{(i)}_j u^{(i)}_j(2) = (1 - \alpha^{(i)}_j) \alpha^{(i)}_j \beta^{(i)}_j u^{(i)}_j(1) + \alpha^{(i)}_j \beta^{(i)}_j u^{(i)}_j(2)\\
    &z^{(i)}_j(3) = (1 - \alpha^{(i)}_j)z^{(i)}_j(2) + \alpha^{(i)}_j \beta^{(i)}_j u^{(i)}_j(3)\\
    &\hphantom{z^{(i)}_j(3)} \quad= {(1 - \alpha^{(i)}_j)}^2 \alpha^{(i)}_j \beta^{(i)}_j u^{(i)}_j(1) + (1 - \alpha^{(i)}_j) \alpha^{(i)}_j \beta^{(i)}_j u^{(i)}_j(2) + \alpha^{(i)}_j \beta^{(i)}_j u^{(i)}_j(3)\\
    &\vdots\\
    &z^{(i)}_j(n) = (1 - \alpha^{(i)}_j)z^{(i)}_j(n - 1) + \alpha^{(i)}_j \beta^{(i)}_j u^{(i)}_j(n) = \sum_{t = 1}^{n} {(1 - \alpha^{(i)}_j)}^{n - t}\alpha^{(i)}_j \beta^{(i)}_j u^{(i)}_j(t)
\end{align*}
for all $j = 1, \ldots, m$.
Hence, the subvalue functions $z^{(i)}(t)$ for all $t = 1, \ldots, n$ can be expressed as
\begin{equation*}
    z^{(i)}(t) = \diag \left(
        \left[
        \begin{array}{cccc}
            \alpha^{(i)}_1\beta^{(i)}_1 & {(1 - \alpha^{(i)}_1)}^1\alpha^{(i)}_1\beta^{(i)}_1 & \cdots & {(1 - \alpha^{(i)}_1)}^{n - 1}\alpha^{(i)}_1\beta^{(i)}_1\\
            \vdots & \vdots & & \vdots\\
            \alpha^{(i)}_m\beta^{(i)}_m & {(1 - \alpha^{(i)}_m)}^1\alpha^{(i)}_m\beta^{(i)}_m & \cdots & {(1 - \alpha^{(i)}_m)}^{n - 1}\alpha^{(i)}_m\beta^{(i)}_m\\
        \end{array}
        \right] \tilde{U}^{(i)}(t)\right),
\end{equation*}
where the matrices $\tilde{U}^{(i)}(t)$ are defined as
\begin{equation}\label{eq:tilde_U_t}
        \tilde{U}^{(i)}(t) = \left[
        \begin{array}{c}
            U^{(i)}(t) \\ 0
        \end{array}
    \right] \in \reals^{n \times m},\quad
    U^{(i)}(t) = \left[
        \begin{array}{c}
            {u^{(i)}(t)}^T\\
            \vdots\\
            {u^{(i)}(1)}^T\\
        \end{array}
    \right] \in \reals^{t \times m}.
\end{equation}
Define the transformation $F \colon \reals^{m} \times \reals^m \to \reals^{m \times n}$, given by
\begin{equation}\label{eq:F}
    F \colon (a, b) \mapsto \left[
        \begin{array}{cccc}
            a_1 b_1 & {(1 - a_1)}^1 a_1 b_1 & \cdots & {(1 - a_1)}^{n - 1} a_1 b_1\\
            \vdots & \vdots & & \vdots\\
            a_m b_m & {(1 - a_m)}^1 a_m b_m & \cdots & {(1 - a_m)}^{n - 1} a_m b_m\\
        \end{array}
        \right],\quad a, b \in \reals^m,
\end{equation}
then the problem (\ref{prob:rl_fit}) can be written as
\begin{equation}\label{prob:rl_fit_F}
    \begin{array}{ll}
    \mbox{minimize} & -\sum_{t=1}^{n} \ell(x(t), y(t))\\[5pt]
    \mbox{subject to} & x(t) = \left[
    \begin{array}{ccc}
        z^{(1)}(t) & \cdots & z^{(k)}(t)
    \end{array}\right] w\\[5pt]
    & z^{(i)}(t) = \diag (F(\alpha^{(i)}, \beta^{(i)}) \tilde{U}^{(i)}(t))\\
    & z^{(i)}(0) = 0,\quad 0 \preceq \alpha^{(i)} \preceq 1,\quad \beta^{(i)} \succeq 0\\
    & i = 1, \ldots, k,\quad t = 1, \ldots, n,
    \end{array}
\end{equation}
where the variables are $\alpha^{(i)}, \beta^{(i)} \in \reals^m$, $z^{(i)}(1), \ldots, z^{(i)}(n) \in \reals^m$ for all $i = 1, \ldots, k$, and $x(1), \ldots, x(n) \in \reals^m$.
The problem data of (\ref{prob:rl_fit_F}) are $w \in \reals^k$, $y(1), \ldots, y(n) \in \reals^m$, and $\tilde{U}^{(i)}(1), \ldots, \tilde{U}^{(i)}(n) \in \reals^{m \times n}$ for all $i = 1, \ldots, k$ given by (\ref{eq:tilde_U_t}).

We can now easily check the convexity of (\ref{prob:rl_fit}) via the equivalent form (\ref{prob:rl_fit_F}).
Note that the objective function in (\ref{prob:rl_fit_F}) can be written as
\begin{equation}\label{eq:rl_fit_obj}
\begin{aligned}
    -\sum_{t=1}^{n} \ell(x(t), y(t)) &= -\sum_{t=1}^{n} \log \left(\frac{{y(t)}^T \exp(x(t))}{\sum_{i=1}^m \exp(x_i(t))}\right)\\
    &= -\sum_{t = 1}^{n} \left({y(t)}^T x(t) - \log\sum_{i = 1}^{m} \exp(x_i(t))\right),
\end{aligned}
\end{equation}
where the second equality is from the fact that, by (\ref{eq:y_t}), the vector $y(t)$ is a standard basis vector for all $t = 1, \ldots, n$.
Since in the last expression of (\ref{eq:rl_fit_obj}), the ${y(t)}^T x(t)$ is affine and the second log-sum-exp term is convex~\cite[\S3.1]{boyd2004convex}, we conclude that the objective of (\ref{prob:rl_fit_F}) is convex in the variables $x(t)$.
(This follows from basic convex analysis~\cite{rockafellar1970convex,boyd2004convex}.)
Then we immediately see that the problem (\ref{prob:rl_fit_F}) is convex if and only if the equality constraints are all affine and the inequality constraints are all convex.
However, this condition is violated by the second constraint
\begin{equation*}
    z^{(i)}(t) = \diag (F(\alpha^{(i)}, \beta^{(i)}) \tilde{U}^{(i)}(t)),\quad i = 1, \ldots, k,\quad t = 1, \ldots, n,
\end{equation*}
since the transformation $F$ given by (\ref{eq:F}) is \emph{not} affine.
Hence, we conclude that the RL model fitting problem (\ref{prob:rl_fit_F}) is \emph{not} convex.
As a result, even if the time complexity of evaluating the objective and constraints of (\ref{prob:rl_fit}) is only $O(mnk)$, the complexity of \emph{solving} it to global optimality in the worst case can be exponential in $mnk$~\cite{nesterov2004introductory,houska2019global}.

\section{Solution method}\label{sec:solution_cvx}
\subsection{The convex surrogate}
Analysis in \S\ref{sec:cvx_alys} indicates that by relaxing the transformation $F$ given by (\ref{eq:F}) to be affine, the RL model fitting problem can then be convexified.
To make such relaxation more explicit, we consider an equivalent formulation of (\ref{prob:rl_fit_F}) given as follows.
For all $i = 1, \ldots, k$, let
\begin{equation*}
    \eta^{(i)} \in \reals^m \quad \mbox{and} \quad
    G^{(i)} = \left[
        \begin{array}{ccc}
            g^{(i)}_1 & \cdots & g^{(i)}_n
        \end{array}
    \right] \in \reals^{m \times n},
\end{equation*}
where $g^{(i)}_1, \ldots, g^{(i)}_n \in \reals^m$ are the columns of the matrix $G^{(i)}$.
The problem (\ref{prob:rl_fit_F}) is then equivalent to
\begin{equation}\label{prob:rl_fit_G}
    \begin{array}{ll}
        \mbox{minimize} & -\sum_{t=1}^{n} \ell(x(t), y(t))\\[5pt]
        \mbox{subject to} & x(t) = \left[
        \begin{array}{ccc}
            z^{(1)}(t) & \cdots & z^{(k)}(t)
        \end{array}\right] w\\[5pt]
        & z^{(i)}(t) = \diag (G^{(i)} \tilde{U}^{(i)}(t)),\quad z^{(i)}(0) = 0\\
        & g^{(i)}_{j+1} = \diag(\eta^{(i)}) g^{(i)}_{j},\quad 0 \preceq \eta^{(i)} \preceq 1,\quad g^{(i)}_n \succeq 0\\
        & i = 1, \ldots, k,\quad j = 1, \ldots, n - 1,\quad t = 1, \ldots, n,
    \end{array}
\end{equation}
where the variables are $\eta^{(i)} \in \reals^m$, $G^{(i)} \in \reals^{m \times n}$, $z^{(i)}(1), \ldots, z^{(i)}(n) \in \reals^m$, $i = 1, \ldots, k$, and $x(1), \ldots, x(n) \in \reals^m$.
The problem data of (\ref{prob:rl_fit_G}) are $w \in \reals^k$, $y(1), \ldots, y(n) \in \reals^m$, and $\tilde{U}^{(i)}(1), \ldots, \tilde{U}^{(i)}(n) \in \reals^{m \times n}$ for all $i = 1, \ldots, k$ given by (\ref{eq:tilde_U_t}).

The equivalence between the formulations (\ref{prob:rl_fit_F}) and (\ref{prob:rl_fit_G}) can be easily verified:
Notice that the constraints
\begin{equation}\label{eq:ncvx_constrs}
    g^{(i)}_{j+1} = \diag(\eta^{(i)}) g^{(i)}_{j},\quad 0 \preceq \eta^{(i)} \preceq 1,\quad g^{(i)}_n \succeq 0
\end{equation}
for all $i = 1, \ldots, k$ and $j = 1, \ldots, n - 1$ essentially enforce the columns of the matrix $G^{(i)}$ in (\ref{prob:rl_fit_G}) to be nonnegative and decay \emph{geometrically} with factor $\eta^{(i)}$ along each respective row.
This is exactly the structure of the transformation $F$ given by (\ref{eq:F}).
In particular, the vectors $\eta^{(i)}$ and $g^{(i)}_1$ in (\ref{prob:rl_fit_G}) correspond to $(1 - \alpha^{(i)})$ and $\diag(\alpha^{(i)}) \diag(\beta^{(i)})$ in (\ref{eq:F}), respectively.

Now we can see that the nonconvexity of (\ref{prob:rl_fit_G}) follows directly from the first equality constraint in (\ref{eq:ncvx_constrs}).
Therefore, to convexify (\ref{prob:rl_fit_G}), we relax the constraints in (\ref{eq:ncvx_constrs}) to
\begin{equation*}
    g^{(i)}_1 \succeq \cdots \succeq g^{(i)}_n,\quad g^{(i)}_n \succeq 0
\end{equation*}
and remove the variables $\eta^{(i)}$ for all $i = 1, \ldots, k$.
This relaxation can be interpreted as simply requiring the entries of the matrices $G^{(i)}$ to decay along each respective row, but not necessarily geometrically.
The resulting relaxed problem
\begin{equation}\label{prob:rl_fit_cvx}
        \begin{array}{ll}
        \mbox{minimize} & -\sum_{t=1}^{n} \ell(x(t), y(t))\\[5pt]
        \mbox{subject to} & x(t) = \left[
        \begin{array}{ccc}
            z^{(1)}(t) & \cdots & z^{(k)}(t)
        \end{array}\right] w\\[5pt]
        & z^{(i)}(t) = \diag (G^{(i)} \tilde{U}^{(i)}(t))\\
        & z^{(i)}(0) = 0,\quad g^{(i)}_1 \succeq \cdots \succeq g^{(i)}_n,\quad g^{(i)}_n \succeq 0\\
        & i = 1, \ldots, k,\quad t = 1, \ldots, n
    \end{array}
\end{equation}
with variables $G^{(i)} \in \reals^{m \times n}$, $z^{(i)}(1), \ldots, z^{(i)}(n) \in \reals^m$, $i = 1, \ldots, k$, and $x(1), \ldots, x(n) \in \reals^m$, is now a convex optimization problem since the objective is convex and the inequality and equality constraints are all affine.
We can then solve (\ref{prob:rl_fit_cvx}) efficiently in many ways, \eg, via interior-point methods~\cite{nesterov1994interior,nocedal1999numerical,boyd2004convex}.
We should note that the time complexity of \emph{evaluating} (\ref{prob:rl_fit_cvx}) is $O(mn^2k)$, which is higher than that of evaluating (\ref{prob:rl_fit}) by a factor of $n$.
However, since (\ref{prob:rl_fit_cvx}) is convex, it can be solved to global optimality in polynomial time.
Specifically, the time complexity of \emph{solving} (\ref{prob:rl_fit_cvx}) is $O(mn^2k)$ if a first-order method is used, and $O({(mn^2k)}^3)$ if a second-order method is used.
In both cases, the time complexity of solving (\ref{prob:rl_fit_cvx}) is expected to be less than that of solving the nonconvex problem (\ref{prob:rl_fit}), which can be exponential in $mnk$ in the worse case.
See \S\ref{sec:expt} for some numerical results on the solution time of (\ref{prob:rl_fit}) and (\ref{prob:rl_fit_cvx}) in applications.

Solving the relaxed problem (\ref{prob:rl_fit_cvx}) as a convex surrogate to the RL model fitting problem (\ref{prob:rl_fit_F}) (or equivalently to (\ref{prob:rl_fit})) has several useful properties.
Let $\alpha^{(i)}$ and $\beta^{(i)}$, $i = 1, \ldots, k$, be in the feasible set of (\ref{prob:rl_fit_F}).
Since (\ref{prob:rl_fit_cvx}) is a relaxation of (\ref{prob:rl_fit_F}), then there must exist feasible point $G^{(i)}$, $i = 1, \ldots, k$, to the problem (\ref{prob:rl_fit_cvx}), such that $G^{(i)} = F(\alpha^{(i)}, \beta^{(i)})$.
Hence, the optimal value of the relaxed problem (\ref{prob:rl_fit_cvx}) gives a lower bound on the optimal value of the RL model fitting problem (\ref{prob:rl_fit_F}).
In particular, suppose the problem (\ref{prob:rl_fit_cvx}) attains its optimum at $G^{(i)\star}$, $i = 1, \ldots, k$.
If there exist $\eta^{(i)} \in \reals^m$ such that $0 \preceq \eta^{(i)} \preceq 1$ and $g^{(i)\star}_{j+1} = \diag(\eta^{(i)}) g^{(i)\star}_{j}$ (where $g^{(i)\star}_{j}$ denotes the $j$th column of $G^{(i)\star}$) for all $i = 1, \ldots, k$ and $j = 1, \ldots, n - 1$, \ie, the constraints (\ref{eq:ncvx_constrs}) are satisfied, then such a lower bound to (\ref{prob:rl_fit_F}) obtained from solving (\ref{prob:rl_fit_cvx}) is tight.
In general, of course, this does not happen --- at least some rows of $G^{(i)\star}$ do not decrease geometrically.
For these cases, we could not say much regarding the tightness of such a lower bound since it is then dependent on the problem data (at least partially).
Nevertheless, numerical examples show that fitting an RL model via (\ref{prob:rl_fit_cvx}) has very similar performance to via (\ref{prob:rl_fit_F}), but has the advantage of tractability.
This can be partially explained as follows.
When the RL model fitting problem is `hard', for example, when the subject's behavior is quite stochastic so the noise levels in the data are high, no fitting method (and, in particular, neither (\ref{prob:rl_fit_F}) nor (\ref{prob:rl_fit_cvx})) can do a good job at recovering the targeted variables.
When the estimation problem is `easy', for example, when the subject's behavior is close to deterministic so the noise levels are low, even simple estimation methods (including via (\ref{prob:rl_fit_cvx})) can do a good job at estimating the (sub)value functions and the RL model parameters.
Therefore it is only problems in between `hard' and `easy' where we could possibly see a significant difference in fitting performance between (\ref{prob:rl_fit_F}) and (\ref{prob:rl_fit_cvx}).
In this region, however, we observe from numerical experiments that they achieve very similar performance.

\subsection{Recovering RL model parameters}\label{sec:param_rec}
Let $G^{(i)\star}$, $i = 1, \ldots, k$, be the optimal point of the relaxed RL model fitting problem (\ref{prob:rl_fit_cvx}).
It is then sufficient for most applications to compute the corresponding (sub)value functions, \ie, $x^\star(t)$ and $z^{(i)\star}(t)$, which are the variables of most interest to researchers.
However, it is sometimes still required to recover the full set of RL model parameters.
Informally, we consider this step as finding a group of feasible $\alpha^{(i)\star}$ and $\beta^{(i)\star}$, such that the difference between the matrices $F(\alpha^{(i)\star}, \beta^{(i)\star})$ and $G^{(i)\star}$ for all $i = 1, \ldots, k$ is minimized.
Note that if the resulting $\alpha^{(i)\star}$ and $\beta^{(i)\star}$ satisfy $F(\alpha^{(i)\star}, \beta^{(i)\star}) = G^{(i)\star}$ for all $i = 1, \ldots, k$, we can conclude that $\alpha^{(i)\star}$ and $\beta^{(i)\star}$ are the (globally) optimal point to (\ref{prob:rl_fit_F}), although, again, this does not happen in general.

The process described above can be formulated mathematically as follows:
Let $\bar{g}^{(i)\star}_j \in \reals^n$ be the vector consisting of the $j$th row of the matrix $G^{(i)\star} \in \reals^{m \times n}$, $i = 1, \ldots, k$, then the $j$th entry to the vectors $\alpha^{(i)\star}, \beta^{(i)\star} \in \reals^m$ can be recovered by solving the problem
\begin{equation}\label{prob:param_rec}
    \begin{array}{ll}
        \mbox{minimize} & {\left\| f(\alpha_j^{(i)}, \beta_j^{(i)}) - \bar{g}_j^{(i)\star} \right\|}_2^2\\
        \mbox{subject to} & 0 \leq \alpha_j^{(i)} \leq 1,\quad \beta_j^{(i)} \geq 0
    \end{array}
\end{equation}
individually for all $i = 1, \ldots, k$, $j = 1, \ldots, m$.
The problem (\ref{prob:param_rec}) has variable $\alpha_j^{(i)}, \beta_j^{(i)} \in \reals$ and data $\bar{g}^{(i)\star} \in \reals^n$, and the transformation $f \colon \reals \times \reals \to \reals^n$ is given by
\begin{equation*}
    f \colon (a, b) \mapsto (ab,\ {(1 - a)}^1 ab,\ \cdots,\ {(1 - a)}^{n - 1} ab),\quad a, b \in \reals.
\end{equation*}
Using a similar argumentation as in \S\ref{sec:cvx_alys}, we may see that the problem (\ref{prob:param_rec}) is \emph{not} convex, and hence, we consider finding a solution to (\ref{prob:param_rec}) via local minimization with repeated initialization.
In other words, we find several local minima of (\ref{prob:param_rec}) from different initial points, and select the one with the least objective value.

One may notice that by choosing a different penalty function for the difference between $f(\alpha_j^{(i)}, \beta_j^{(i)})$ and $\bar{g}_j^{(i)\star}$ in (\ref{prob:param_rec}), the problem of recovering the RL model parameters can be formulated as a convex program.
We leave the corresponding discussion for this approach in \S\ref{sec:param_rec_cvx} of the supplementary material, and will not consider it further in this paper for the reasons listed there.

\subsection{Truncation of the horizon}\label{sec:trunc_horizon}
Notice that for all time steps $t = 1, \ldots, n$, we can approximate the calculation of each entry of the subvalue function $z^{(i)}(t)$ as
\begin{equation}\label{eq:trunc}
    z^{(i)}_j(t) = \sum_{\tau = 1}^{t} {(1 - \alpha^{(i)}_j)}^{t - \tau}\alpha^{(i)}_j \beta^{(i)}_j u^{(i)}_j(\tau)
    \approx \sum_{\tau = t - p + 1}^{t} {(1 - \alpha^{(i)}_j)}^{t - \tau}\alpha^{(i)}_j \beta^{(i)}_j u^{(i)}_j(\tau),
\end{equation}
for all $i = 1, \ldots, k$, $j = 1, \ldots, m$, since the term ${(1 - \alpha^{(i)}_j)}^{t - \tau}$ can be very close to zero for small $\tau$.
The approximation (\ref{eq:trunc}) can be interpreted as truncating the horizon to the last $p$ steps when accumulating the subreward signals, instead of using the full history until the start of the episode.
That is, the current subvalue function $z^{(i)}(t)$ is only dependent on the last $p$ subreward signals $u^{(i)}(t - p + 1), \ldots, u^{(i)}(t)$ (zero padding when $\tau \leq 0$).
As two extreme examples, if $p = n$, no truncation is applied;
if $p = 1$, the subvalue function $z^{(i)}(t)$ can be determined only from $u^{(i)}(t)$, \ie, there is no ``memory'' in the decision process.

The approximation (\ref{eq:trunc}) can be easily integrated into the RL model fitting problem (\ref{prob:rl_fit_F}) by replacing the transformation $F$ defined in (\ref{eq:F}) with $F_p \colon \reals^m \times \reals^m \to \reals^{m \times p}$ given by
\begin{equation*}
    F_p \colon (a, b) \mapsto \left[
        \begin{array}{cccc}
            a_1 b_1 & {(1 - a_1)}^1 a_1 b_1 & \cdots & {(1 - a_1)}^{p - 1} a_1 b_1\\
            \vdots & \vdots & & \vdots\\
            a_m b_m & {(1 - a_m)}^1 a_m b_m & \cdots & {(1 - a_m)}^{p - 1} a_m b_m\\
        \end{array}
        \right],\quad a, b \in \reals^m.
\end{equation*}
Correspondingly, we replace $\tilde{U}^{(i)}(t)$ defined by (\ref{eq:tilde_U_t}) with the submatrices $\tilde{U}_p^{(i)}(t) \in \reals^{m \times p}$, which consist of the first $p$ rows of $\tilde{U}^{(i)}(t)$ for all $t = 1, \ldots, n$.
Finally, the relaxed problem (\ref{prob:rl_fit_G}) and the problem (\ref{prob:param_rec}) for recovering RL model parameters can be easily adapted.

In practice, the horizon length $p$ is a hyperparameter chosen by the user a priori.
(See \S\ref{sec:select_p} for some discussion on how to select the value of $p$ in practice.)
When $n$ is large and $p \ll n$, introducing the approximation (\ref{eq:trunc}) can significantly decrease the solving time, since the number of (scalar) variables in (\ref{prob:rl_fit_G}) is reduced from $kmn$ to $kmp$.
This corresponds to a reduction of the time complexity of evaluating (\ref{prob:rl_fit_cvx}) from $O(mn^2k)$ to $O(m p^2 k)$.

\section{Implementation}\label{sec:impl}
In this section we describe our implementation of the ideas described in \S\ref{sec:solution_cvx} for fitting RL model to behavioral data under multi-armed bandits.
The source code has been collated into an open-source Python package \texttt{rlfit}, which is freely available online at
\begin{quote}
    \url{https://github.com/nrgrp/rlfit}.
\end{quote}

The core module in the \texttt{rlfit} package is the \texttt{RLFit} class.
At initialization, \texttt{RLFit} takes an integer and a boolean to specify the horizon length $p$ (see \S\ref{sec:trunc_horizon}) and whether the model parameters are shared across bandits (as described in \S\ref{sec:forget_qlearning}), respectively.
To fit the RL model to some data via solving the relaxed problem (\ref{prob:rl_fit_cvx}), the user calls the \texttt{fit} method.
This method implements a solver for (\ref{prob:rl_fit_cvx}) based on the domain specific language \texttt{CVXPY}~\cite{diamond2016cvxpy,agrawal2018rewriting} for convex optimization problems, and takes mainly the following arguments:
\begin{itemize}
    \item \texttt{rewards}: A \texttt{numpy} array that has the shape \texttt{(n, m)} or a list of such \texttt{numpy} arrays (with each array representing a subreward signal), corresponding to the data $u^{(i)}(t) \in \reals^m$, $i = 1, \ldots, k$, $t = 1, \ldots, n$, in (\ref{prob:rl_fit}).
    \item \texttt{actions}: A \texttt{numpy} array with shape \texttt{(n, m)}, corresponding to the problem data $y(t) \in \{e_1, \ldots, e_m\} \subseteq \reals^m$, $t = 1, \ldots, n$, in (\ref{prob:rl_fit}).
    \item \texttt{w}: A number or a \texttt{numpy} array with shape \texttt{(k,)}, corresponding to the data $w \in \reals^k$ in (\ref{prob:rl_fit}).
        If a number is given, it is automatically transformed into a $k$-dimensional \texttt{numpy} array by repeating the same given number $k$ times.
\end{itemize}
Then, if the user would like to recover the RL model parameters $\alpha^{(i)\star}$ and $\beta^{(i)\star}$, the method \texttt{fit\_param} will be called subsequently.
This method finds a solution to the problem (\ref{prob:param_rec}) via repeated local minimization using \texttt{SciPy}~\cite{virtanen2020scipy}.
Note that the argument \texttt{concurrent} for this method is set to \texttt{True} by default.
This allows the problem (\ref{prob:param_rec}) with different data, \ie, individual rows of $G^{(i) \star}$, to be minimized in parallel on multiple CPU cores.
As a result, all entries of the parameter vectors $\alpha^{(i)\star}$ and $\beta^{(i)\star}$ can be recovered semi-simultaneously.

Once the RL model is fit, it can be used via either the \texttt{predict} or \texttt{score} method.
The two arguments of \texttt{predict} are the data \texttt{rewards} and \texttt{w}, as in the \texttt{fit} method.
This function returns the predicted action selection probability for all time steps, according to (\ref{eq:fql_pi}), as well as the corresponding underlying (sub)value functions $x^\star(t)$ and $z^{(i)\star}(t)$, $i = 1, \ldots, k$, $t = 1, \ldots, n$.
The \texttt{score} method takes the same first three arguments as the \texttt{fit} method, and evaluates the log-likelihood of the dataset, \ie, the negative objective of the problem (\ref{prob:rl_fit}).
Note that the only prerequisite for using the \texttt{predict} and \texttt{score} method is to call the \texttt{fit} method during model fitting, and calling the \texttt{fit\_param} method is not mandatory.
If the RL model is fit only via the \texttt{fit} method, the functions implemented in \texttt{predict} and \texttt{score} will be based on the optimal point $G^{(i)\star}$, $i = 1, \ldots, k$, for the relaxed problem (\ref{prob:rl_fit_cvx});
if both \texttt{fit} and \texttt{fit\_param} are called, the methods \texttt{predict} and \texttt{score} will instead use $\alpha^{(i)\star}$ and $\beta^{(i)\star}$, $i = 1, \ldots, k$, from the problem (\ref{prob:param_rec}).

\section{Empirical experiments}\label{sec:expt}
In this section, we evaluate our method proposed in \S\ref{sec:solution_cvx} for fitting RL models under several popular multi-armed bandit environments.
We compare our approach with two other solution methods that appear most commonly in the literature.

\subsection{Environment setup}\label{sec:env_setup}
We consider the following three multi-armed bandit environment setups, with each assigned a three capital letter tag which we will refer to during subsequent discussion.
\begin{itemize}
    \item \textsf{BSC}: The basic bandit environment defined according to the basic forgetting Q-learning model, given by (\ref{eq:fql_rewsig}) to (\ref{eq:fql_pi}).
        The model fitting problem corresponds to (\ref{prob:fql}).
    \item \textsf{IND}: Extend the \textsf{BSC} setup by incorporating different learning rate $\alpha$ and reward sensitivity $\beta$ for individual choices.
        The model fitting problem corresponds to (\ref{prob:multi_lr}).
    \item \textsf{SUB}: Extend the \textsf{IND} setup by further incorporating two subreward signals and subvalue functions.
        The first subreward signal is the same as the reward signal used for \textsf{BSC} and \textsf{IND}, given by (\ref{eq:fql_rewsig}).
        The second subreward signal is equal to $y(t)$ given by (\ref{eq:y_t}) for all $t = 1, \ldots, n$.
        (This setup is sometimes considered to model the subject's behavior of repeating the last action under bandit tasks~\cite{beron2022mice}.)
        The coefficient $w$ used to combine the subvalue functions is defined as the most general case, \ie, $w = \ones$.
        Then the model fitting problem corresponds to (\ref{prob:subrew}) with $k = 2$.
\end{itemize}
Each of the three setups consists of a smaller (2-armed, $m = 2$) and a larger (10-armed, $m = 10$) version.
The smaller version has a reward probability $(0.9, 0.1)$ for each possible action, and after each action selection, there is a $0.02$ chance of shuffling the reward probabilities.
Similarly, the larger version has the reward probability
\[
    (0.30, 0.27, 0.95, 0.67, 0.69, 0.29, 0.42, 0.05, 0.73, 1.00)
\]
for each action, but there is no reward shuffling.
The 2-armed bandit environment targets at simulating the animal behavior experiment task widely used for rodents, \eg, in~\cite{hattori2019area,hamaguchi2022prospective}, while the larger version aims at those tasks designed for human, \eg, in~\cite{kleespies2025sleep}.
Although even the larger version `only' consists of 10 arms, it indeed covers almost all environments that appear in real-world behavioral experiments.
For simplicity in description, we assign the tag `\textsf{2AB}' to the 2-armed bandit environment and `\textsf{10AB}' to the 10-armed setup.

For each environment setup, we collected a dataset consisting of $1000$ episodes, where each episode has $200$ time steps (\ie, $n = 200$).
For each episode, the model parameters $\alpha$ and $\beta$ were randomly sampled from a uniform distribution defined on the intervals (or boxes) given by table~\ref{tab:range_param}.

\begin{table}[h]
    \centering
    \caption{Range of model parameters.}\label{tab:range_param}
    \begin{tabular}{@{}cccc@{}}
    \toprule
        & \textsf{BSC} & \textsf{IND} & \textsf{SUB} \\ \midrule
    \textsf{2AB}  & $\alpha \in [0, 1]$, $\beta \in [0, 5]$   & $\alpha \in {[0, 1]}^2$, $\beta \in {[0, 5]}^2$   & \makecell{$\alpha^{(1)} \in {[0, 1]}^2$, $\beta^{(1)} \in {[0, 5]}^2$\\$\alpha^{(2)} \in {[0, 1]}^2$, $\beta^{(2)} \in {[0, 2]}^2$}   \\[15pt]
    \textsf{10AB} & $\alpha \in [0, 1]$, $\beta \in [5, 10]$   & $\alpha \in {[0, 1]}^{10}$, $\beta \in {[5, 10]}^{10}$   & \makecell{$\alpha^{(1)} \in {[0, 1]}^{10}$, $\beta^{(1)} \in {[5, 10]}^{10}$\\$\alpha^{(2)} \in {[0, 1]}^{10}$, $\beta^{(2)} \in {[0, 5]}^{10}$}   \\ \bottomrule
    \end{tabular}
\end{table}

\subsection{Benchmarks}\label{sec:benchmarks}
\subsubsection{Direct local minimization with repeated initiation}
As shown in \S\ref{sec:cvx_alys}, fitting an RL model to behavioral data consists in solving some instance of the nonconvex optimization problem (\ref{prob:rl_fit}).
In practice, one of the most direct approaches is to just apply local minimization methods repeatedly from different initial points~\cite{beron2022mice}.
Then, the locally optimal point with the best performance (in the case of (\ref{prob:rl_fit}), this corresponds to the least cumulative negative log-likelihood value) is returned as the final (approximate) solution.
Although there are a huge range of local minimization algorithms, one should note that directly minimizing (\ref{prob:rl_fit}) needs to include bound constraints on the model parameters $\alpha$ and $\beta$.
For this purpose, the following solvers are widely considered and easily accessible via the Python library \texttt{SciPy}~\cite{virtanen2020scipy}:
Nelder-Mead~\cite{nelder1965simplex,gao2012implementing}, L-BFGS-B~\cite{zhu1997algorithm}, TNC~\cite{nash1984newton}, SLSQP~\cite{kraft1988software}, Powell~\cite{powell1964efficient}, trust region with constraints (Trust-Region)~\cite{branch1999subspace,conn2000trust}, COBYLA~\cite{powell1994direct}, and COBYQA~\cite{powell2002uobyqa,ragonneau2022modelbased,ragonneau2024cobyqa}.
Note that the Nelder-Mead algorithm is a simplex method that does not support constraints inherently, but simply handles the box constraints by just clipping all vertices in the simplex based on the bounds.
All the aforementioned solvers are evaluated in our empirical experiments (see \S\ref{sec:results} and supplementary tables~\ref{tab:basic_fql_2arm} to~\ref{tab:subrew_10arm} for numerical results).

\subsubsection{Bayesian modeling}
Another popular approach for fitting RL models to behavioral data is to use Bayesian modeling.
This method aims at estimating the posterior distribution of the model parameters given the observed subject actions $y(t)$ for all $t = 1, \ldots, n$.
Then, the parameters corresponding to the highest posterior probability are selected as an approximate solution to the fitting problem.

With slight abuse of notation, let $\alpha = (\alpha^{(1)}, \ldots, \alpha^{(k)}) \in \reals^{mk}$ and $\beta = (\beta^{(1)}, \ldots, \beta^{(k)}) \in \reals^{mk}$, then the target distribution of the Bayesian estimation process is written as
\begin{equation}\label{eq:map_target}
    p(\alpha, \beta \mid y(1), \ldots, y(n)) \propto p(y(1), \ldots, y(n) \mid \alpha, \beta) p(\alpha, \beta).
\end{equation}
In general, the prior distribution $p(\alpha, \beta) = p(\alpha) p(\beta)$ is given by uniform distributions on the feasible set for the respective parameters, \ie,
\begin{equation*}
    p(\alpha) = \uni(0, \ones),\quad p(\beta) = \uni(0, \beta_{\rm max}),
\end{equation*}
where $\beta_{\rm max} \in \reals^{mk}_{++}$ is the prior information for an upper bound on $\beta$.
Under these uniform priors, roughly speaking, the goal of the Bayesian estimation process is equivalent to solving the RL model fitting problem (\ref{prob:rl_fit}) with an additional box constraint $\beta \preceq \beta_{\rm max}$~\cite[\S7.2]{boyd2004convex}.
Although one may consider some distribution with support $[0, \infty)$ to make $p(\beta)$ consistent with the corresponding constraints in (\ref{prob:rl_fit}), it is then difficult to choose the parameters that controls the shape of the priors.
Besides, it may take more effort to accurately estimate the target posterior.
Nevertheless, it has been reported empirically in the literature that the Bayesian modeling approach for fitting RL models is not very sensitive to the choice of the prior distribution~\cite{hamaguchi2022prospective,kleespies2025sleep}.
Finally, after obtaining the posterior $p(\alpha, \beta \mid y(1), \ldots, y(n))$, one may choose the parameters $\alpha^\star$ and $\beta^\star$ corresponding to the highest density as a solution to (\ref{prob:rl_fit}).

In practice, the posterior distribution $p(\alpha, \beta \mid y(1), \ldots, y(n))$ given in (\ref{eq:map_target}) is often estimated via Monte Carlo methods.
This approach for fitting RL models is considered in several studies~\cite{hamaguchi2022prospective,kleespies2025sleep}, but it is generally less used in practice than direct local minimization methods.
This is probably due to the difficulties in the implementation and debugging of a Monte Carlo method solver, even though the Python library \texttt{PyMC}~\cite{abril2023pymc} has significantly simplified this procedure.
On the other hand, since the posterior distribution of the model parameters provides some global information about the solution space of (\ref{prob:rl_fit}), RL model fitting via Bayesian modeling is expected to be more robust to the local minima issue than direct local minimization methods.

\subsection{Solver configurations}\label{sec:solver_config}
In this section, we list the configuration details for our solution method introduced in \S\ref{sec:solution_cvx}, as well as the two benchmarks introduced in \S\ref{sec:benchmarks}.
When evaluating each solver, the RL model fitting was performed individually for each episode collected under all environments.
The name tags used in the subsequent discussion and the configurations corresponding to each solver are listed as follows:
\begin{itemize}
    \item \textsf{MC}: Bayesian estimation via Monte Carlo.
        The prior distributions for model parameters are set as uniform distributions on the range given in table~\ref{tab:range_param}.
        For the fitting of all episodes, the number of sampled Markov chains was set to $4$, with each consisting of $2000$ burn-in samples and $5000$ estimation samples.
    \item \textsf{D-LOC}: Directly solve the model fitting problem (\ref{prob:rl_fit}) via local minimization methods with repeated initializations.
        For the fitting of each episode under different environments, the initial values of the model parameters (\ie, the optimization variables) were sampled from a uniform distribution on the range given in table~\ref{tab:range_param}.
        The number of repeated initializations is set to $5$.
    \item \textsf{CVX}: Solve the corresponding convex relaxation problem (\ref{prob:rl_fit_cvx}).
        Note that this approach does not recover the RL model parameters $\alpha^{(i)\star}$ and $\beta^{(i)\star}$, $i = 1, \ldots, k$.
        Hence, the evaluation was only performed based on the estimated value functions $x^\star(t)$, $t = 1, \ldots, n$.
    \item \textsf{CVX-T}: The same as \textsf{CVX}, but with a truncated horizon (\ref{eq:trunc}), where $p = 5$.
    \item \textsf{CVX-LOC}: First perform \textsf{CVX} and then recover the model parameters by finding a solution to (\ref{prob:param_rec}) via local minimization methods.
        For the second local minimization step, when fit for each episode under different environments, the initial values of the model parameters were sampled from a uniform distribution on the range given in table~\ref{tab:range_param}.
        The number of repeated initializations is again set to $5$.
        Note that the concurrent computing feature for the parameter recovery problem (\ref{prob:param_rec}) with different data as introduced in \S\ref{sec:impl} is enabled.
    \item \textsf{CVX-LOC-T}: The same as \textsf{CVX-LOC}, but with the truncated horizon approximation (\ref{eq:trunc}), where $p = 5$, for the first \textsf{CVX} step.
\end{itemize}
Note that to make the prior information compatible across different methods (in particular, between \textsf{MC} and the the other methods), in the numerical experiments, we adapted the constraints about $\beta^{(i)}$ in (\ref{prob:rl_fit}) from $\beta^{(i)} \succeq 0$ to $\beta^{(i)}_{\rm min} \preceq \beta^{(i)} \preceq \beta^{(i)}_{\rm max}$ for all $i = 1, \ldots, k$.
The bounds $\beta^{(i)}_{\rm min}$ and $\beta^{(i)}_{\rm max}$ are defined according to table~\ref{tab:range_param}.
In addition, when a local minimization step is required, all algorithms listed in the first paragraph of \S\ref{sec:benchmarks} were applied individually.

\subsection{Evaluation metrics}\label{sec:metrics}
To evaluate the performance of different solution methods, we consider the following two metrics.
Firstly, the performance of recovering the value functions $x(t)$, $t = 1, \ldots, n$, is commonly measured by an indirect metric --- the KL-divergence of the corresponding true and recovered action selection probability.
Specifically, let
\begin{equation*}
    \pi(t) = \frac{\exp(x(t))}{\sum_{i = 1}^{m} \exp(x_i(t))} \in \reals^m,\quad t = 1, \ldots, n,
\end{equation*}
which is the probability of selecting individual actions at the $t$th time step.
Then for each episode, we calculate the mean KL-divergence between the ground truth $\pi^{\rm gt}(t)$ and the estimated $\pi^\star(t)$ (obtained using the ground truth $x^{\rm gt}(t)$ and the recovered $x^\star(t)$ respectively), across $t = 1, \ldots, n$, \ie,
\begin{equation*}
    \expect D_{\rm kl}(\pi^{\rm gt}(t), \pi^\star(t)) = \frac{1}{n}\sum_{t = 1}^{n} D_{\rm kl}(\pi^{\rm gt}(t), \pi^\star(t)).
\end{equation*}
Secondly, to measure the error of the recovered model parameters, we use $\norm{\alpha^{\rm gt} - \alpha^\star}_2$ and $\norm{\beta^{\rm gt} - \beta^\star}_2$, where $\alpha^{\rm gt}$ and $\beta^{\rm gt}$ are the ground truth model parameters, and $\alpha^\star$ and $\beta^\star$ are the corresponding estimations.
Note that here (as well as in the subsequent discussion) the notation $\alpha$ and $\beta$ can refer to real numbers (for \textsf{BSC} setup), or vectors in $\reals^m$ (for \textsf{IND} setup), or even the concatenated real vectors (for \textsf{SUB} setup), given by $\alpha = (\alpha^{(1)}, \ldots, \alpha^{(k)}) \in \reals^{mk}$, $\beta = (\beta^{(1)}, \ldots, \beta^{(k)}) \in \reals^{mk}$.
The exact meaning of these notation can be determined from the context (or the text).
For the \textsf{BSC} setup, such metric is simply the absolute value of the difference.

\subsection{Numerical results}\label{sec:results}
\begin{figure}[p]
    \centering
    \begin{subfigure}[b]{\textwidth}
        \centering
        \includegraphics[width=\textwidth]{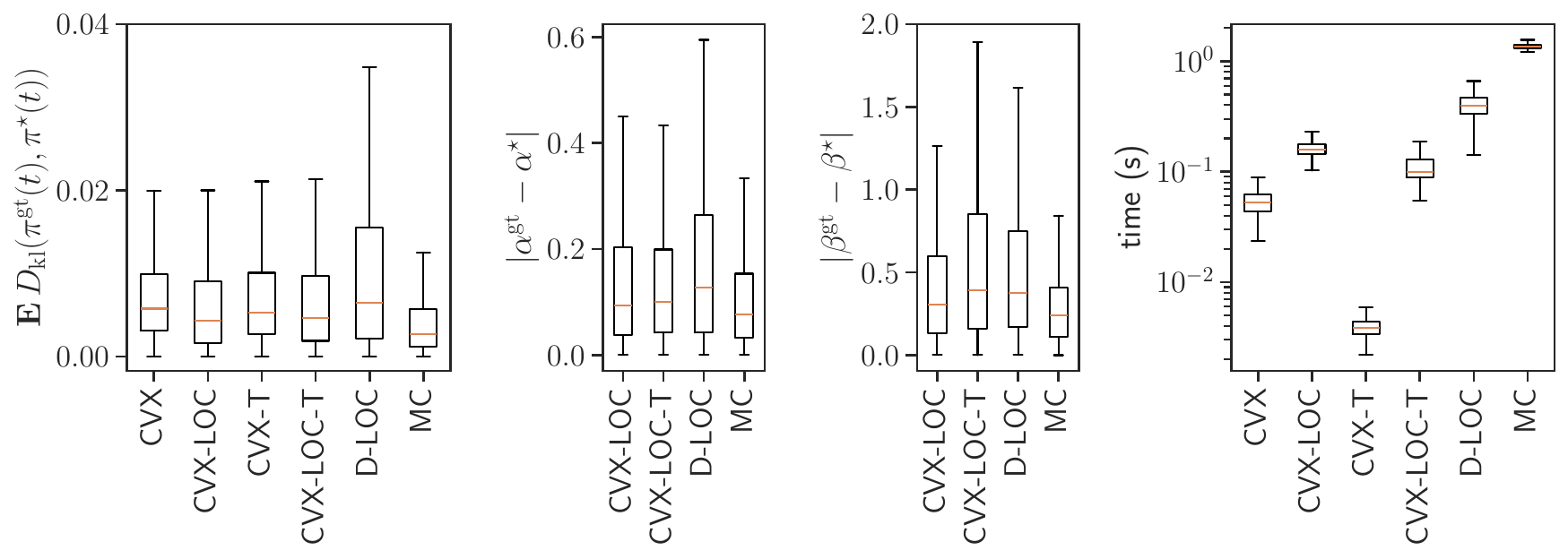}
        \caption{The \textsf{BSC} setup.}\label{sfig:basic_fql_2arm}
    \end{subfigure}\\[10pt]
    \begin{subfigure}[b]{\textwidth}
        \centering
        \includegraphics[width=\textwidth]{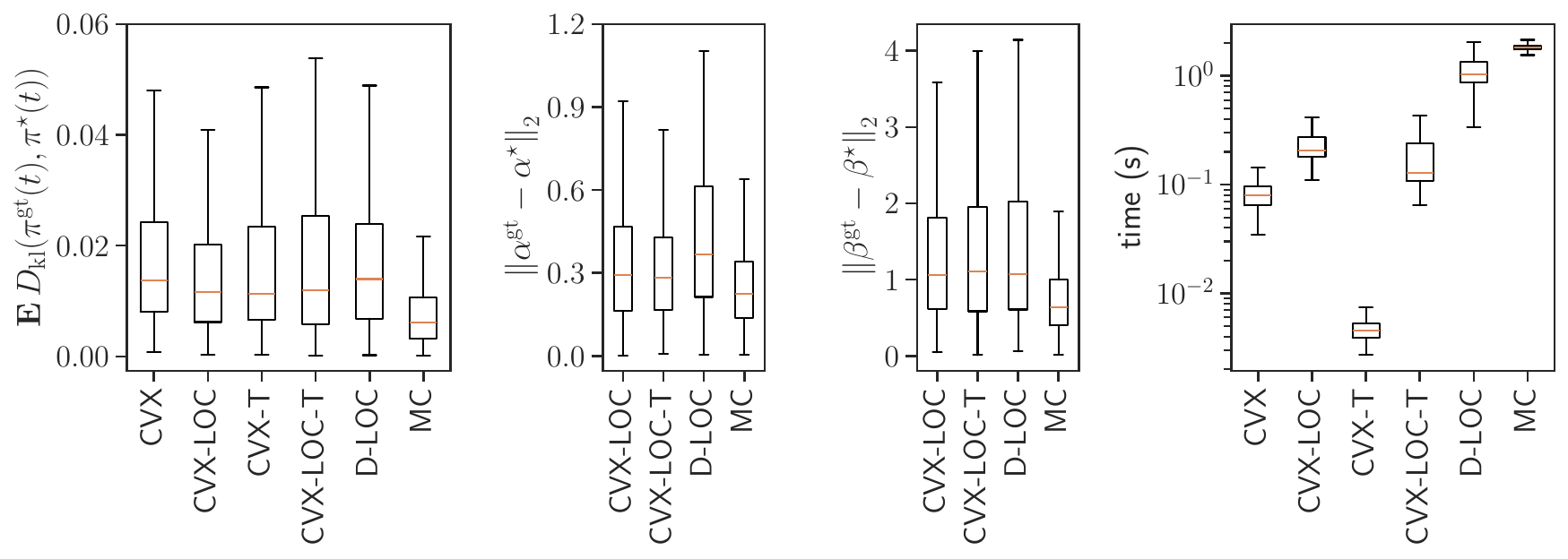}
        \caption{The \textsf{IND} setup.}\label{sfig:multi_lr_2arm}
    \end{subfigure}\\[10pt]
    \begin{subfigure}[b]{\textwidth}
        \centering
        \includegraphics[width=\textwidth]{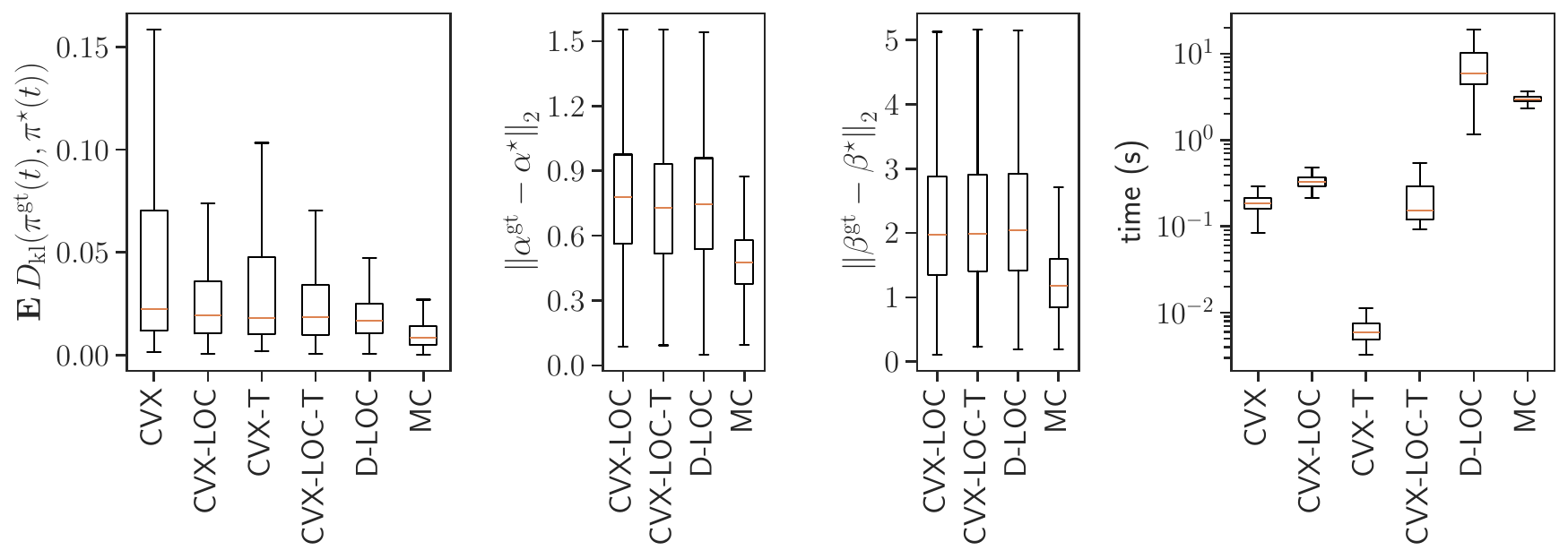}
        \caption{The \textsf{SUB} setup.}\label{sfig:subrew_2arm}
    \end{subfigure}\\[10pt]
    \caption{
        Performance of different solution methods for fitting RL models in the \textsf{2AB} environment.
        The left column shows the mean KL-divergence between the ground truth and recovered action selection probability; the middle column shows the error of the recovered model parameters; and the right column shows the computing time required for fitting the RL model.
    }\label{fig:2ab}
\end{figure}

\begin{figure}[p]
    \centering
    \begin{subfigure}[b]{\textwidth}
        \centering
        \includegraphics[width=\textwidth]{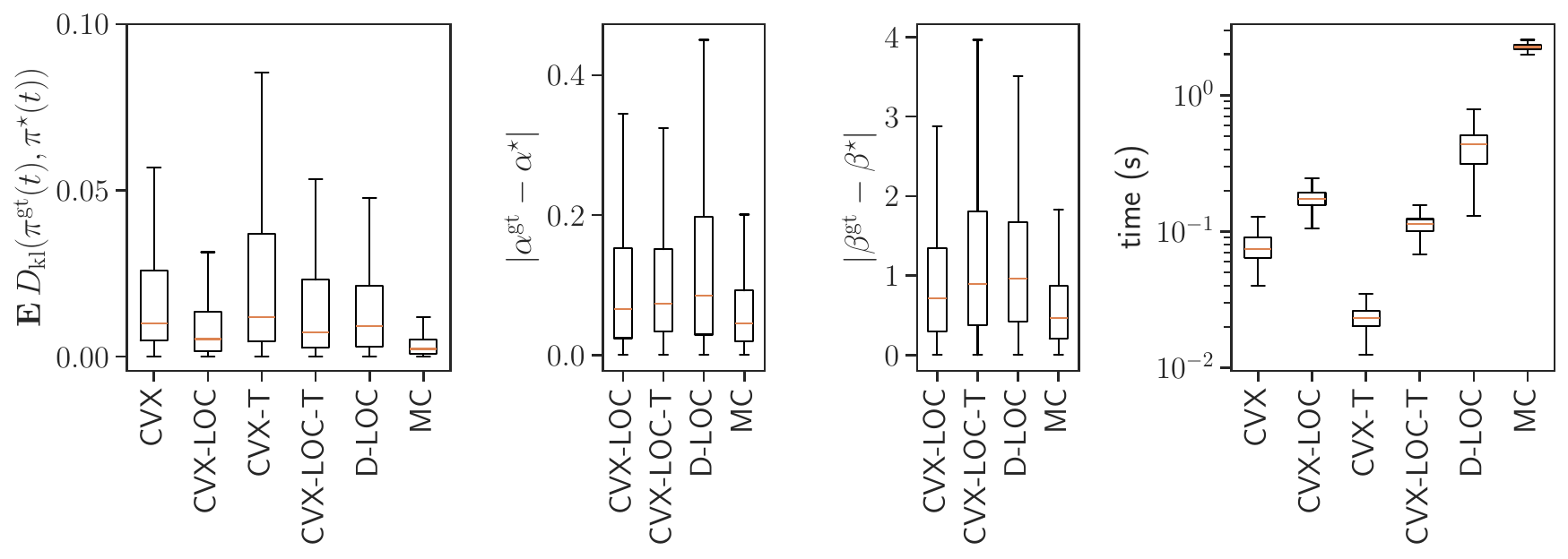}
        \caption{The \textsf{BSC} setup.}\label{sfig:basic_fql_10arm}
    \end{subfigure}\\[10pt]
    \begin{subfigure}[b]{\textwidth}
        \centering
        \includegraphics[width=\textwidth]{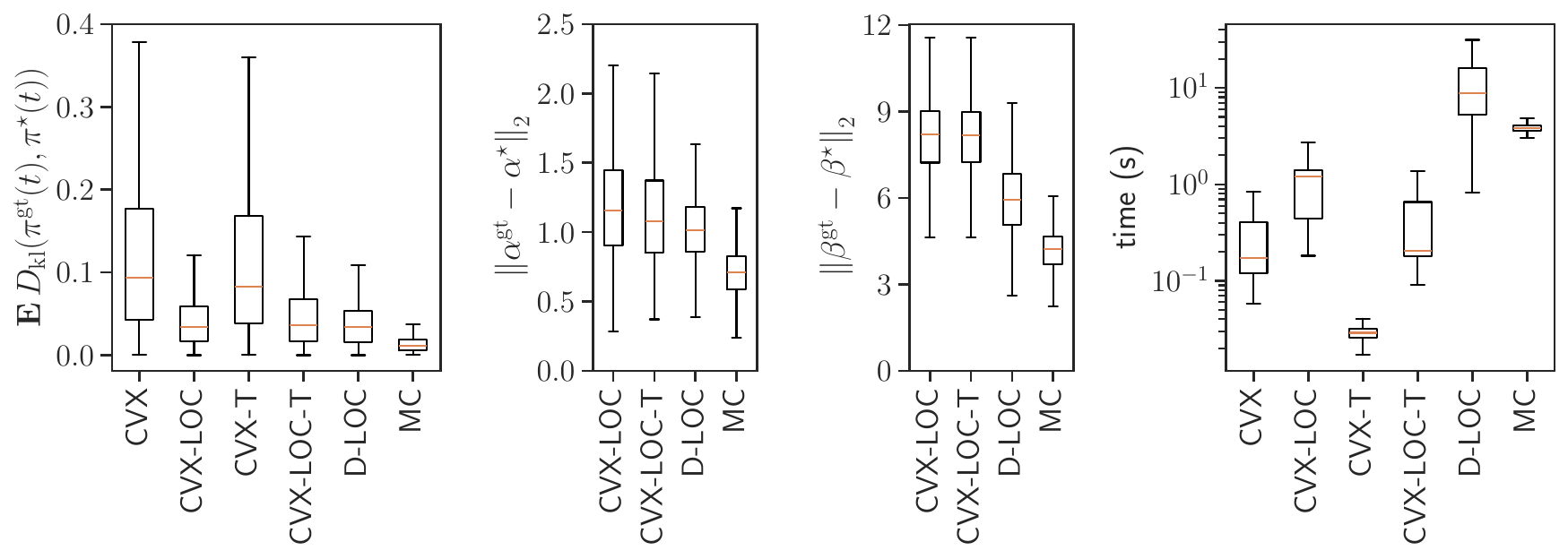}
        \caption{The \textsf{IND} setup.}\label{sfig:multi_lr_10arm}
    \end{subfigure}\\[10pt]
    \begin{subfigure}[b]{\textwidth}
        \centering
        \includegraphics[width=\textwidth]{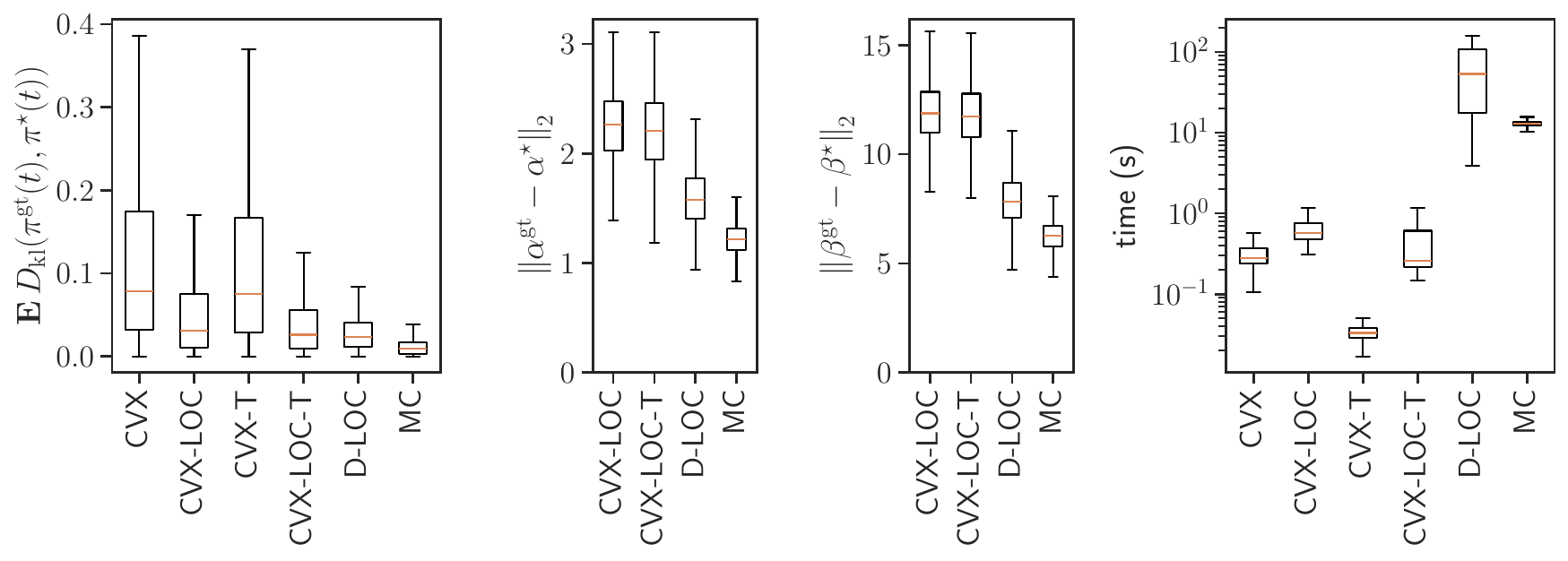}
        \caption{The \textsf{SUB} setup.}\label{sfig:subrew_10arm}
    \end{subfigure}\\[10pt]
    \caption{
        Performance of different solution methods for fitting RL models in the \textsf{10AB} environment.
        The left column shows the mean KL-divergence between the ground truth and recovered action selection probability; the middle column shows the error of the recovered model parameters; and the right column shows the computing time required for fitting the RL model.
    }\label{fig:10ab}
\end{figure}

According to our experiments, the major difference between different local minimization solvers as listed in the first paragraph of \S\ref{sec:benchmarks} only appears in computing time (as shown in supplementary tables~\ref{tab:basic_fql_2arm}--\ref{tab:subrew_10arm}).
Therefore, we select the Trust-Region algorithm as the default solver in the following discussion (figures~\ref{fig:2ab} and~\ref{fig:10ab}) because of its robustness, numerical stability, and broad applicability across different problem scales.
Readers may refer to \S\ref{sec:app_table} of the supplementary material for the detailed numerical values corresponding to the figures~\ref{fig:2ab} and~\ref{fig:10ab}, as well as those results from the other local minimization solvers that are not included in the figures.

\paragraph{The \textsf{2AB} environment.}
In general, under the \textsf{2AB} environment and across all three setups (\textsf{BSC}, \textsf{IND}, and \textsf{SUB}), as expected, the \textsf{MC} method had the best performance both in recovering the value functions and the model parameters.
All other methods had similar performance but slightly below \textsf{MC} (figure~\ref{fig:2ab}, columns 1--3).
Surprisingly, the additional truncated horizon approximation (\ref{eq:trunc}) in \textsf{CVX-T} and \textsf{CVX-LOC-T} did not result in a significant loss of solution accuracy.
In terms of computational efficiency, our convex surrogate based methods, \textsf{CVX}, \textsf{CVX-LOC}, \textsf{CVX-T}, and \textsf{CVX-LOC-T}, had faster computing time compared to \textsf{D-LOC} and \textsf{MC} methods across all setups.
Specifically, under the \textsf{BSC} setup, the \textsf{CVX-T} method had the fastest solving time within $10^{-2}$ second;
\textsf{CVX}, \textsf{CVX-LOC}, and \textsf{CVX-LOC-T} were slightly slower at the level of $10^{-1}$ second, while \textsf{D-LOC} and \textsf{MC} required even more computing time, for approximately $4 \times 10^{-1}$ and $1.4$ second(s), respectively.
As the setup got more complicated from \textsf{BSC} to \textsf{SUB}, the required solving time also increased for all methods.
In particular, the \textsf{D-LOC} method took even longer (in median) than \textsf{MC} for solving the fitting problem under the \textsf{SUB} setup (figure~\ref{fig:2ab}, last column).
This indicates that the \textsf{D-LOC} method has the highest sensitivity to the RL model scale, whereas our convex surrogate based methods are the least influenced ones (which is a direct result of the concurrent implementation as introduced in \S\ref{sec:impl}).

\paragraph{The \textsf{10AB} environment.}
The performance of different solution methods in terms of the fitting accuracy under the \textsf{10AB} environment (figure~\ref{fig:10ab}) is more or less similar to those from the \textsf{2AB} environment.
The slight differences only appear under the \textsf{IND} and \textsf{SUB} setups (figures~\ref{sfig:multi_lr_10arm} and~\ref{sfig:subrew_10arm}).
Specifically, the accuracy of recovering the subjects' action selection probability via \textsf{CVX} and \textsf{CVX-T}, measured by the mean KL-divergence $\expect D_{\rm kl}(\pi^{\rm gt}(t), \pi^\star(t))$, has a larger median value and variance compared to the other methods.
However, such a minor decrease does not have much influence in practice, since the real-world environments that subjects encounter are unlikely to be as complicated.
One may notice that under the most complex setup \textsf{SUB}, our convex surrogate based method also resulted in a larger fitting error regarding the accuracy of recovering the RL model parameters (figure~\ref{sfig:subrew_10arm}, two middle columns).
On the one hand, since the accuracy level about the subject's action selection probability does not vary much across different solution methods, it is reasonably expected that there exist multiple groups of RL model parameters that could lead to the same observed behavior.
On the other hand, noticing that in this case the vectors for evaluating $\norm{\alpha^{\rm gt} - \alpha^\star}_2$ and $\norm{\beta^{\rm gt} - \beta^\star}_2$ are all in $\reals^{20}$, the estimation error for each parameter is hence still below $1$, which does not really influence the real-world applications either.
The results for the computing time of different solution methods under the \textsf{10AB} environment are almost exactly the same as those observed under \textsf{2AB}, except that the solving time increased significantly for all methods (figure~\ref{fig:10ab}, last column).
Notably, with setups \textsf{IND} and \textsf{SUB}, the \textsf{D-LOC} method may take much longer than the \textsf{MC} method, but result in even worse performance in recovering the value functions and model parameters.

\section{Example}
This section provides an example of using our provided \texttt{rlfit} package for fitting RL models to real-world behavioral data under bandits.
We also compare our methods with the two benchmarks introduced in \S\ref{sec:benchmarks} in terms of the fitting performance and computing time.

\subsection{The dataset}
We consider the \emph{two-armed bandit reversal learning task}, which is one of the standard protocols widely used in animal behavioral neuroscience research~\cite{hattori2019area,hamaguchi2022prospective,de2023freibox,zhu2024multiintention,beron2022mice}.
In particular, the dataset used here is from the previous work by De La Crompe~\etal~\cite{de2023freibox} and Zhu~\etal~\cite{zhu2024multiintention}.

The dataset consists of the behavior of $9$ different mice performing the reversal learning task, with a total of $64$ sessions (where each session corresponds to one episode in our previous discussion).
In this task, the animal subject is presented with two water spouts, and it can select one of the spouts to receive a water reward.
At the beginning of each session, a random spout is assigned a water reward and the other spout is not rewarded, and the animal may freely select either of the two spouts at each time step.
As the session proceeds, the animal needs to learn which spout is currently rewarded.
If the animal was able to collect $75\%$ of the rewards in the last $15$ time steps, the reward contingency will be reversed, \ie, the previously unrewarded spout becomes rewarded and the previously rewarded spout becomes unrewarded.
After each reversal, the animal needs to adapt its behavior to the new reward contingency.
In this dataset, each session consists of $3$ reversals, \ie, $4$ blocks with different reward contingencies in total, and the total number of time steps in each session is roughly $100$.

\subsection{Models and fitting}
We consider fitting three different RL models under the setup assumptions \textsf{BSC}, \textsf{IND}, and \textsf{SUB} as presented in \S\ref{sec:env_setup} to the given dataset, and compare the final results between the solution methods \textsf{MC}, \textsf{D-LOC}, \textsf{CVX-T}, and \textsf{CVX-LOC-T} as listed in \S\ref{sec:solver_config}.

\subsubsection{Selecting the horizon length}\label{sec:select_p}
The solution methods \textsf{CVX-T} and \textsf{CVX-LOC-T} require the user to specify a horizon length $p$ for the truncated horizon approximation (\ref{eq:trunc}).
There are several ways to select the value of $p$ in practice.
Empirically, it has been reported that a horizon length of $3$ to $5$ is often sufficient for fitting RL models to behavioral data under such tasks~\cite{hattori2019area, zhu2024multiintention,beron2022mice}.
Quantitatively (or at least semi-quantitatively), one can also solve the problem (\ref{prob:rl_fit_cvx}) with approximation (\ref{eq:trunc}) under different horizon length $p$, and select the one that achieves the best balance between the optimal value of the RL model fitting problem (\ie, the best log-likelihood of the observed data) and the total number of problem variables.

\begin{figure}
    \centering
    \includegraphics[width=0.5\textwidth]{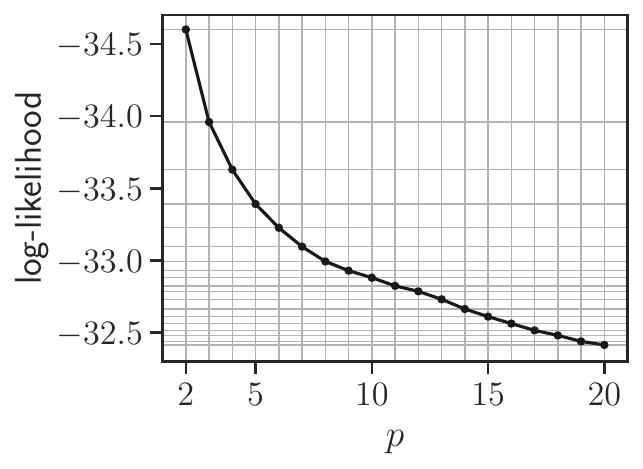}
    \caption{
        Log-likelihood values of the observed data, obtained from solving the convex surrogate (\ref{prob:rl_fit_cvx}) with approximation (\ref{eq:trunc}), under different horizon lengths $p$.
        Each point in the figure corresponds to the mean log-likelihood across all episodes in the dataset (with standard error approximately $1.9$, not shown).
    }\label{fig:select_p}
\end{figure}

Here we consider the second approach as an example to select the horizon length $p$.
We apply the method \textsf{CVX-T} with different horizon lengths $p$ varying from $2$ to $20$ to fit the RL model under the \textsf{BSC} setup to the dataset.
Then, the fitting performance is evaluated by the log-likelihood of the observed data, \ie, the negative of the objective of the problem (\ref{prob:rl_fit_cvx}).
The results are shown in figure~\ref{fig:select_p}.
It is observed that, in general, the fitting performance gets better as the horizon length $p$ grows, but the improvement becomes less significant when $p$ is larger than $5$.
This is consistent with the empirical observations in the literature, and hence we may select $p = 5$ as a reasonable value for the subsequent data analysis.

\subsubsection{Model fitting}
The following code snippet shows how to fit the RL model under the three setups using our package \texttt{rlfit}.
For each session in the dataset, suppose the data is loaded into the arrays \texttt{rews} and \texttt{acts}, to fit the RL model under the \textsf{BSC} setup via \textsf{CVX-T}, we can use the following code:
\begin{lstlisting}[language=zhpython]
import rlfit as rf
# define the BSC model with horizon length p
model = rf.RLFit(horizon_len=p, share_param=True)
# fit the model to the data
model.fit(rews, acts)
# evaluate the log-likelihood of the data under the fitted model
model.score(rews, acts)
\end{lstlisting}
To fit the RL model under the \textsf{IND} setup, we can simply set the argument \texttt{share\_param} to \texttt{False} when defining the model, and to fit the RL model under the \textsf{SUB} setup, we need to pass \texttt{[rews, acts]} as the first argument to the \texttt{fit} method.
Finally, to analyze the whole dataset, we can just repeat the above procedure for each session.

The code snippet above for the \textsf{CVX-T} method can be easily adapted to the \textsf{CVX-LOC-T} method by simply adding the following code after the \texttt{fit} method:
\begin{lstlisting}[language=zhpython]
# recover the model parameters
model.fit_param(max_beta, number_repeats)
\end{lstlisting}
Here the argument \texttt{max\_beta} is a hyperparameter that defines an upper bound for the model parameter $\beta$, and \texttt{number\_repeats} is the number of repeated initializations for the local minimization solver.
All values of the hyperparameters used in this example (for all applied solution methods) remain the same as those described in \S\ref{sec:solver_config}.
The implementation of the other two benchmarks roughly involves several tens of lines of Python code, and hence is not included here.
We refer interested readers to the companion code repository of this article for the details.

\subsection{Results}\label{sec:eg_result}
\begin{figure}
    \centering
    \includegraphics[width=\textwidth]{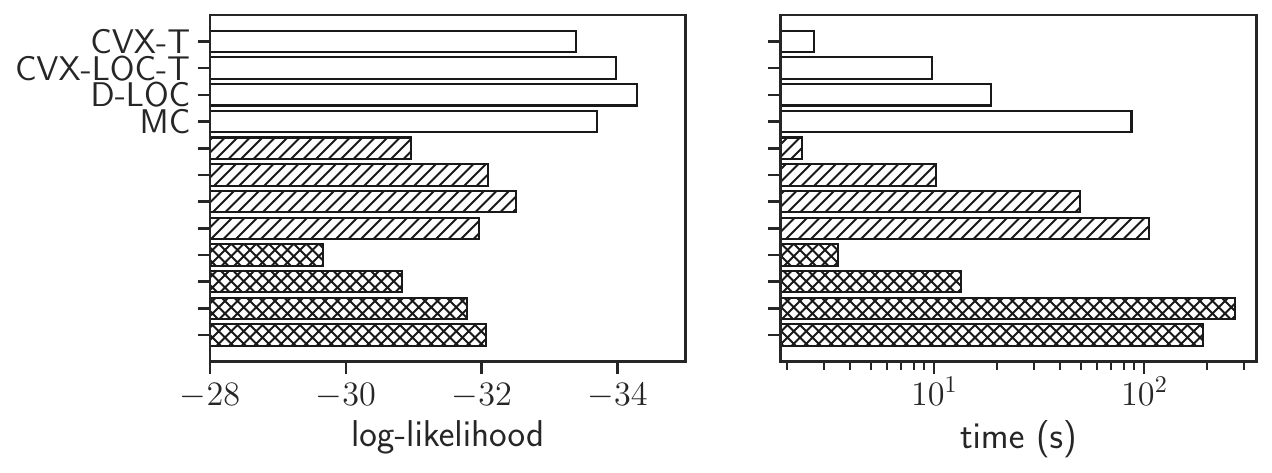}
    \caption{
        \emph{Left.}
        Best model fitting performance in terms of the log-likelihood of the observed data, obtained from different solution methods.
        The empty, line-hatched, and cross-hatched bars correspond to the \textsf{BSC}, \textsf{IND}, and \textsf{SUB} setups, respectively.
        The name of each method is only shown for the \textsf{BSC} setup, but the same order applies to the other two setups as well.
        Each bar in the figure shows the mean log-likelihood across all episodes in the dataset (with standard error approximately $1.9$, $1.8$, and $1.7$ for \textsf{BSC}, \textsf{IND}, and \textsf{SUB}, respectively; not shown).
        \emph{Right.}
        Total computing time required for fitting different RL models to the whole dataset via different solution methods.
    }\label{fig:eg_result}
\end{figure}

The best log-likelihood of the observed data under the fitted models via different solution methods, and the corresponding computing time, is shown in the left and right of figure~\ref{fig:eg_result}, respectively.

For all three setups, the \textsf{CVX-T} method achieves the best fitting performance in terms of the log-likelihood.
This follows from the feature that this method (roughly, because of the approximation (\ref{eq:trunc})) finds a lower bound for the original RL model fitting problem (\ref{prob:rl_fit}) via the convex surrogate (\ref{prob:rl_fit_cvx}).
The corresponding compromise is then not fully recovering the model parameters and slightly violating the RL model assumptions.
After recovering the model parameters via the local minimization step, the \textsf{CVX-LOC-T} method achieves a slightly worse fitting performance than \textsf{CVX-T}.
This aligns with our expectation since the relaxations introduced by \textsf{CVX-T} are removed and the RL model assumptions are fully enforced.
Nevertheless, the fitting performance of \textsf{CVX-LOC-T} is still comparable to the \textsf{MC} and \textsf{D-LOC} methods.

In terms of the computing time, the observations from figure~\ref{fig:eg_result} are more or less the same as those observed in the numerical experiments with synthetic data:
The \textsf{CVX-T} method is the fastest one, followed by \textsf{CVX-LOC-T}, while the \textsf{D-LOC} and \textsf{MC} methods, again, require significantly more computing time.

\section{Conclusion and discussion}
\subsection{Summary of numerical results}
Numerical results as listed in \S\ref{sec:results} and \S\ref{sec:eg_result} suggest that, in general, our convex surrogate based method for fitting RL models to behavioral data under multi-armed bandits achieves comparable performance as the other two benchmarks, but with significantly decreased computing time.
Although the sampling based Bayesian estimation method is most likely to result in the best performance in terms of the fitting accuracy, its sampling process may last quite long.
This disadvantage can be problematic when the behavioral dataset consists of a large number of episodes and the computing time is constrained.
In comparison, our method achieves a good balance between the solution accuracy and the computing time.

Our observations also suggest that, although more as a byproduct, the most commonly used local minimization approach may not be ideal, as it achieves only moderate solution accuracy while requiring comparable or even greater computational time than the other methods.
To the best of our knowledge, the preference for this direct method in the literature may be due to the relative simplicity of implementing and debugging, compared to the Monte Carlo sampling procedures.
Our convex surrogate based method, however, offers a relatively clean and straightforward procedure that can be easily implemented by users familiar with convex optimization.
Moreover, we also provide a generic, well documented Python package implementing our proposed solution method, allowing scientific researchers without prior knowledge about convex optimization to apply it in the analysis of their datasets directly.

\subsection{Judging a heuristic fitting}
Recall that our method, by solving the convex surrogate (\ref{prob:rl_fit_cvx}), computes a lower bound for the original problem (\ref{prob:rl_fit}).
In particular, such a lower bound is computed for each specific problem instance (\ie, RL model structure) and data (\ie, observed subject behavior).
This property can be applied to evaluate the suboptimality of any other heuristic fitting results (at least semi-quantitatively).

Let $J$ be a heuristic objective value of (\ref{prob:rl_fit}) (obtained via any solution method), and let $J^\star$ be the global minimum.
Suppose that by solving the convex program (\ref{prob:rl_fit_cvx}), we obtain a lower bound $J^{\rm lb}$ to (\ref{prob:rl_fit}), then we have the inequalities
\[
    J^{\rm lb} \leq J^\star \leq J.
\]
If $J - J^{\rm lb}$ is small, then we may conclude that such a heuristic solution is nearly (globally) optimal, and the bound $J^{\rm lb}$ is nearly tight.
If $J - J^{\rm lb}$ is big, then for this problem instance and data, either the fitting is poor, or, the bound is poor (or both).

\subsection{Parameter recovery}
One may notice that our convex surrogate based method via the problem (\ref{prob:rl_fit_cvx}) does not directly recover the parameters $\alpha^{(i)}$ and $\beta^{(i)}$ of the original RL model, but only estimates the value functions $x(t)$.
Then the parameter recovery step is performed by solving the problem (\ref{prob:param_rec}), which is a nonconvex optimization problem.
Although such two-step method does not fully resolve the nonconvexity issue in the original RL model fitting problem (\ref{prob:rl_fit}), it still has several advantages compared to directly solving (\ref{prob:rl_fit}).

Firstly, in many practical scenarios, researchers' main interest of fitting RL models to behavioral data is to recover the subjects' value functions~\cite{hattori2019area,hattori2023meta}.
This can be directly obtained from a solution of the convex surrogate (\ref{prob:rl_fit_cvx}), without the need for the next step of parameter recovery.
According to our numerical results, the accuracy of recovering the value functions from the convex surrogate in the most commonly used two-armed bandit settings is already comparable to the other two benchmarks (figures~\ref{fig:2ab}, first column), but the computational cost is significantly reduced.
Secondly, the RL model parameters recovery step via (\ref{prob:param_rec}) allows parallel computing of $\alpha^{(i)}$ and $\beta^{(i)}$ for different $i = 1, \ldots, k$.
In contrast, directly solving (\ref{prob:rl_fit}) requires all model parameters to be optimized together.
Empirical results suggest that this concurrent feature of our method significantly improves computational efficiency, especially when the scale of the environment gets larger (figures~\ref{fig:2ab} and~\ref{fig:10ab}, last column).

To sum up, the major contribution of our proposed method is not in finding a better (approximate) solution to the original nonconvex RL model fitting problem that is more close to the global optimum, but in providing a computationally efficient approach that can be applied to larger scale model setups and datasets.

\subsection{Previous and related work}
\subsubsection{Generalized linear models}
Readers that are familiar with generalized linear models might notice that our proposed solution method for fitting RL models can be interpreted as transforming the RL model into a multi-label logistic regression model, whose fitting problem is well known to be a convex optimization problem.
A similar connection between these two types of models was previously observed and discussed by Beron~\etal~\cite{beron2022mice}, although their focus was primarily on comparing different models of behavior under bandit settings, particularly in terms of interpretability and how well various behavioral characteristics were captured.
Based on empirical data and observations, they argued that the generalized logistic regression model and the original RL model are approximately equivalent.
In contrast, our convex analysis in this paper offers a deeper theoretical insight, showing that the generalized logistic regression model is, in fact, a convex relaxation of the original RL model.

\subsubsection{Inverse reinforcement learning}
Inverse RL is a well-known problem in the field of machine learning, which aims at recovering the reward function or/and the policy of an agent based on its observed behavior~\cite{ng2000algorithms,abbeel2004apprenticeship}.
It is closely related to the problem of fitting RL models to behavioral data, as both problems involve inferring the underlying decision making process of an agent from its demonstrations.
Specifically, the problem considered in this work can be seen as a special case of inverse RL, applied to the setting of multi-armed bandits, where the state space and transition dynamics are trivial.

Inverse RL has been a very popular research topic in the machine learning community, and has a wide range of applications in robotics, autonomous driving, and human-computer interaction.
We refer the interested readers to Shao and Er~\cite{zhifei2012survey,zhifei2012review}, Arora and Doshi~\cite{arora2021survey}, and Adams~\etal~\cite{adams2022survey} for some reviews on the recent developments and applications.
Moreover, inverse RL has also been applied in characterizing animal and human behavior in neuroscience and psychology research~\cite{kalweit2022neurl,alyahyay2023mechanisms,ruiz2023inverse}.
One of the most notable recent advances in this direction is the \emph{hierarchical} or \emph{multi-intention} inverse RL framework, where the agent's behavior is assumed to be generated by multiple latent intentions or subgoals.
These approaches have so far provided many novel explanations for the observed subject behavior in complex environments~\cite{zhu2024multiintention,ke2025inverse,wang2025learning,shehab2025learning,hausladen2026identifying}.
Besides, hierarchical inverse RL has also attained much attention in the fields of, \eg, engineering~\cite{feng2025deep} and medicine~\cite{kalweit2025ai,vogt2025ai}.
On the other hand, the model fitting problem of such hierarchical models is often very challenging, since it often involves solving a hierarchical optimization problem, where the inner and outer loop problems correspond to inverse RL and the latent intention estimation, respectively.
The proposed method in this work can be used as a computationally efficient building block for these hierarchical models on multi-armed bandits.
In particular, via the convex relaxation (\ref{prob:rl_fit_cvx}), the inner loop inverse RL problem of these hierarchical models can be solved efficiently.
Most importantly, these ideas can be implemented and prototyped via the recent developed multi-convex programming frameworks~\cite{shen2017disciplined,zhu2025disciplined}, and is hence easily actionable by practitioners.

\subsubsection{Bandits and the inverse problems}
The current major research focus on bandits is on the design of algorithms for solving the forward problem, \ie, how to find an optimal policy for a given bandit environment.
Some good textbooks and reviews on this topic include those by Bubeck and Cesa-Bianchi~\cite{bubeck2012regret}, Slivkins~\cite{slivkins2019introduction}, Lattimore and Szepesv\'ari~\cite{lattimore2020bandit}, and Zhou~\etal~\cite{zhou2025selective}; just to mention a few.
The inverse problem of bandits, in contrast, has received much less attention and is often considered as a special case of inverse RL.
To the best of our knowledge, the only work that has directly considered the inverse problem of multi-armed bandits is from H{\"u}y{\"u}k~\etal~\cite{huyuk2022inverse}.
Hence, we believe that our work can be seen as a complementary contribution to this line of research.

\newpage
\section*{Acknowledgments}
This work has been supported by BrainLinks-BrainTools, funded by the Ministry of Science, Research and the Arts Baden-W\"urttemberg within the sustainability program for projects of the Excellence Initiative~II; by CRC/TRR~384 ``IN-CODE''; and by the Deutsche Forschungsgemeinschaft (DFG, German Research Foundation) - Project-ID~499552394 - SFB~1597.

\section*{Data availability}
The source code to generate the data used in our numerical experiments and reproduce the results can be found at \url{https://github.com/nrgrp/fit_rl_mab}.

\newpage
\appendix
\section{A convex formulation for recovering RL model parameters}\label{sec:param_rec_cvx}
In this section, we discuss an option of formulating the optimization problem of recovering the RL model parameters from the optimal point $G^{(i)\star}$ of the problem (\ref{prob:rl_fit_cvx}), $i = 1, \ldots, k$, as a convex program.

Instead of directly penalizing the difference between $f(\alpha_j^{(i)}, \beta_j^{(i)})$ and $\bar{g}_j^{(i)\star}$ using the $\ell_2$-squared penalty function as in (\ref{prob:param_rec}), we consider measuring the difference between these two terms in the log space, \ie, for all $i = 1, \ldots, k$, $j = 1, \ldots, m$, we solve the following problem:
\begin{equation}\label{prob:param_rec_cvx}
    \begin{array}{ll}
        \mbox{minimize} & {\left\| \log f(\alpha_j^{(i)}, \beta_j^{(i)}) - \log \bar{g}_j^{(i)\star} \right\|}_2^2\\
        \mbox{subject to} & 0 \leq \alpha_j^{(i)} \leq 1,\quad \beta_j^{(i)} \geq 0
    \end{array}
\end{equation}
with optimization variables $\alpha_j^{(i)}, \beta_j^{(i)} \in \reals$ and data $\bar{g}^{(i)\star} \in \reals^n$ from the $j$th row of $G^{(i)\star}$, and the transformation $f \colon \reals \times \reals \to \reals^n$ is given by
\begin{equation*}
    f \colon (a, b) \mapsto (ab,\ {(1 - a)}^1 ab,\ \cdots,\ {(1 - a)}^{n - 1} ab),\quad a, b \in \reals.
\end{equation*}
The objective of the problem (\ref{prob:param_rec_cvx}) can be explicitly written as
\begin{align*}
    {\left\| \log f(\alpha_j^{(i)}, \beta_j^{(i)}) - \log \bar{g}_j^{(i)\star} \right\|}_2^2
    &= {\left\| \left[
        \begin{array}{c}
            \log (\alpha_j^{(i)}\beta_j^{(i)})\\
            \log ({(1 - \alpha_j^{(i)})}^1\alpha_j^{(i)}\beta_j^{(i)})\\
            \vdots\\
            \log ({(1 - \alpha_j^{(i)})}^{n - 1}\alpha_j^{(i)}\beta_j^{(i)})
        \end{array}
    \right] - \log \bar{g}_j^{(i)\star} \right\|}_2^2\\
    &= {\left\| \left[
        \begin{array}{c}
            0 \times \log (1 - \alpha_j^{(i)}) + \log (\alpha_j^{(i)}\beta_j^{(i)})\\
            1 \times \log (1 - \alpha_j^{(i)}) + \log(\alpha_j^{(i)}\beta_j^{(i)})\\
            \vdots\\
            (n - 1) \times \log (1 - \alpha_j^{(i)}) + \log(\alpha_j^{(i)}\beta_j^{(i)})
        \end{array}
    \right] - \log \bar{g}_j^{(i)\star} \right\|}_2^2.
\end{align*}
Introducing the variable transformations
\begin{equation*}
    A = \left[
        \begin{array}{cc}
            0 & 1\\
            1 & 1\\
            \vdots & \vdots\\
            n - 1 & 1
        \end{array}
    \right],\quad
    v_j^{(i)} = \left[
        \begin{array}{c}
            \log (1 - \alpha_j^{(i)}) \\ \log (\alpha_j^{(i)}\beta_j^{(i)})
        \end{array}
    \right],\quad
    s_j^{(i)} = \log \bar{g}_j^{(i)\star},
\end{equation*}
we transform the problem (\ref{prob:param_rec_cvx}) into an equivalent constrained least squares problem
\begin{equation}\label{prob:param_rec_cvx_ls}
    \begin{array}{ll}
        \mbox{minimize} & {\left\|A v_j^{(i)} - s_j^{(i)}\right\|}_2^2\\
        \mbox{subject to} & v_{j,1}^{(i)} < 0
    \end{array}
\end{equation}
with variables $v_j^{(i)} \in \reals^2$ and data $A \in \reals^{n \times 2}$, $s_j^{(i)} \in \reals^n$.
(The notation $v_{j,1}^{(i)}$ denotes the first entry of the vector $v_j^{(i)}$.)

Formulating the problem of recovering the RL model parameters as (\ref{prob:param_rec_cvx}) benefits from the property that it can be solved robustly and efficiently via the convex constrained least squares problem (\ref{prob:param_rec_cvx_ls}).
However, such a formulation may suffer from severe numerical issues.
First, notice that the inequality constraint $v_{j,1}^{(i)} < 0$ of (\ref{prob:param_rec_cvx_ls}) has to be strict, which is not supported by most conic solvers for convex optimization problems.
If we just solve with the relaxed constraint $v_{j,1}^{(i)} \leq 0$, it is likely that we obtain some optimal point $v_{j,1}^{(i)\star}$ with the first entry being $0$, in which case the corresponding optimal point $\alpha_j^{(i)\star} = 0$, where the second entry $\log (\alpha_j^{(i)\star}\beta_j^{(i)\star})$ does not exist for any $\beta_j^{(i)\star} \geq 0$.
Besides, the objective of (\ref{prob:param_rec_cvx}) tends to add a very large penalty to those entries of $\bar{g}_j^{(i)\star}$ that are close to zero compared to the large entries, even when they have the same residual.
Such behavior is due to the characteristic of the logarithmic function $x \mapsto \log x$, whose slope approaches infinity as $x \to 0$.
Recall that in (\ref{prob:rl_fit_cvx}), we expect that each row of the optimal point $G^{(i)\star}$ to decay (ideally) geometrically.
Combining these two effects together, the solution of (\ref{prob:param_rec_cvx}) with data $\bar{g}_j^{(i)}$ tends to have a very good fit to the tail of the vector $\bar{g}_j^{(i)}$ where the entries are nearly zero, while the residual at the beginning entries can be very large.
If $n$ is large, \ie, there might exist many nearly zero entries in $\bar{g}_j^{(i)}$, the results can be problematic since they tend to provide a fitting that matches the noninformative small tails of $\bar{g}_j^{(i)}$ which are largely influenced by numerical roundoff error, instead of the informative entries at the beginning.
For these reasons, although we demonstrate this option of formulating the problem of recovering the RL model parameters as the convex program (\ref{prob:param_rec_cvx}), we will not consider this approach in practice.

\section{Additional tables}\label{sec:app_table}
In tables~\ref{tab:basic_fql_2arm} to~\ref{tab:subrew_10arm}, we list the detailed numerical results of all experiments in \S\ref{sec:expt}, corresponding to figures~\ref{fig:2ab} and~\ref{fig:10ab}.
As introduced in \S\ref{sec:metrics}, the expectation term appearing in the first row of each table is over $t = 1, \ldots, n$.

\begin{sidewaystable}
    \caption{Numerical results correspond to figure~\ref{sfig:basic_fql_2arm}.}\label{tab:basic_fql_2arm}
    \resizebox{\textwidth}{!}{%
    \begin{tabular}{@{}ccccccccccc@{}}
    \toprule
    & & Nelder-Mead & L-BFGS-B & TNC & SLSQP & Powell & Trust-Region & COBYLA & COBYQA & --\\ \midrule
    \multirow{6}{*}[-25pt]{\rotatebox{90}{\small$\expect D_{\rm kl}(\pi^{\rm gt}(t), \pi^\star(t))$}} & \textsf{CVX}    & -- & -- & -- & -- & -- & -- & -- & -- & \makecell{$0.006$\\($0.003$--$0.010$)} \\
    & \textsf{CVX-LOC}    & \makecell{$0.004$\\($0.002$--$0.009$)} & \makecell{$0.004$\\($0.002$--$0.009$)} & \makecell{$0.004$\\($0.002$--$0.009$)} & \makecell{$0.004$\\($0.002$--$0.009$)} & \makecell{$0.004$\\($0.002$--$0.009$)} & \makecell{$0.004$\\($0.002$--$0.009$)} & \makecell{$0.004$\\($0.002$--$0.009$)} & \makecell{$0.004$\\($0.002$--$0.009$)} & -- \\
    & \textsf{CVX-T} & -- & -- & -- & -- & -- & -- & -- & -- & \makecell{$0.005$\\($0.003$--$0.010$)} \\
    & \textsf{CVX-LOC-T} & \makecell{$0.005$\\($0.002$--$0.010$)} & \makecell{$0.005$\\($0.002$--$0.010$)} & \makecell{$0.005$\\($0.002$--$0.010$)} & \makecell{$0.005$\\($0.002$--$0.010$)} & \makecell{$0.005$\\($0.002$--$0.010$)} & \makecell{$0.005$\\($0.002$--$0.010$)} & \makecell{$0.005$\\($0.002$--$0.010$)} & \makecell{$0.005$\\($0.002$--$0.010$)} & -- \\
    & \textsf{D-LOC}    & \makecell{$0.006$\\($0.002$--$0.016$)} & \makecell{$0.007$\\($0.002$--$0.017$)} & \makecell{$0.007$\\($0.002$--$0.016$)} & \makecell{$0.007$\\($0.002$--$0.017$)} & \makecell{$0.007$\\($0.002$--$0.016$)} & \makecell{$0.006$\\($0.002$--$0.016$)} & \makecell{$0.006$\\($0.002$--$0.016$)} & \makecell{$0.007$\\($0.002$--$0.016$)} & -- \\
    & \textsf{MC}     & -- & -- & -- & -- & -- & -- & -- & -- & \makecell{$0.003$\\($0.001$--$0.006$)} \\ \midrule
    \multirow{5}{*}[-9pt]{\rotatebox{90}{$|\alpha^{\rm gt} - \alpha^\star|$}} & \textsf{CVX-LOC} & \makecell{$0.10$\\($0.04$--$0.21$)} & \makecell{$0.09$\\($0.04$--$0.20$)} & \makecell{$0.09$\\($0.04$--$0.20$)} & \makecell{$0.09$\\($0.04$--$0.20$)} & \makecell{$0.09$\\($0.04$--$0.20$)} & \makecell{$0.09$\\($0.04$--$0.20$)} & \makecell{$0.09$\\($0.04$--$0.20$)} & \makecell{$0.09$\\($0.04$--$0.20$)} & -- \\
    & \textsf{CVX-LOC-T} & \makecell{$0.10$\\($0.04$--$0.20$)} & \makecell{$0.10$\\($0.04$--$0.20$)} & \makecell{$0.10$\\($0.04$--$0.20$)} & \makecell{$0.10$\\($0.04$--$0.20$)} & \makecell{$0.10$\\($0.04$--$0.20$)} & \makecell{$0.10$\\($0.04$--$0.20$)} & \makecell{$0.10$\\($0.04$--$0.20$)} & \makecell{$0.10$\\($0.04$--$0.20$)} & -- \\
    & \textsf{D-LOC}    & \makecell{$0.13$\\($0.04$--$0.28$)} & \makecell{$0.13$\\($0.04$--$0.27$)} & \makecell{$0.13$\\($0.04$--$0.28$)} & \makecell{$0.13$\\($0.04$--$0.28$)} & \makecell{$0.13$\\($0.04$--$0.28$)} & \makecell{$0.13$\\($0.04$--$0.26$)} & \makecell{$0.13$\\($0.04$--$0.28$)} & \makecell{$0.13$\\($0.04$--$0.28$)} & -- \\
    & \textsf{MC}     & -- & -- & -- & -- & -- & -- & -- & -- & \makecell{$0.08$\\($0.03$--$0.15$)} \\ \midrule
    \multirow{5}{*}[-9pt]{\rotatebox{90}{$|\beta^{\rm gt} - \beta^\star|$}} & \textsf{CVX-LOC} & \makecell{$0.30$\\($0.13$--$0.58$)} & \makecell{$0.31$\\($0.14$--$0.60$)} & \makecell{$0.30$\\($0.13$--$0.59$)} & \makecell{$0.31$\\($0.13$--$0.60$)} & \makecell{$0.30$\\($0.13$--$0.59$)} & \makecell{$0.31$\\($0.13$--$0.60$)} & \makecell{$0.31$\\($0.14$--$0.59$)} & \makecell{$0.31$\\($0.13$--$0.60$)} & -- \\
    & \textsf{CVX-LOC-T} & \makecell{$0.38$\\($0.16$--$0.82$)} & \makecell{$0.38$\\($0.16$--$0.82$)} & \makecell{$0.38$\\($0.16$--$0.83$)} & \makecell{$0.39$\\($0.16$--$0.83$)} & \makecell{$0.38$\\($0.15$--$0.82$)} & \makecell{$0.39$\\($0.16$--$0.85$)} &\makecell{$0.40$\\($0.16$--$0.81$)} &\makecell{$0.39$\\($0.16$--$0.85$)} & -- \\
    & \textsf{D-LOC}    & \makecell{$0.35$\\($0.17$--$0.72$)} & \makecell{$0.37$\\($0.17$--$0.73$)} & \makecell{$0.37$\\($0.17$--$0.73$)} & \makecell{$0.37$\\($0.17$--$0.73$)} & \makecell{$0.38$\\($0.17$--$0.75$)} & \makecell{$0.38$\\($0.17$--$0.75$)} & \makecell{$0.41$\\($0.18$--$0.79$)} & \makecell{$0.41$\\($0.18$--$0.79$)} & -- \\
    & \textsf{MC}     & -- & -- & -- & -- & -- & -- & -- & -- & \makecell{$0.24$\\($0.11$--$0.41$)} \\ \midrule
    \multirow{6}{*}[-20pt]{\rotatebox{90}{time (ms)}} & \textsf{CVX}    & -- & -- & -- & -- & -- & -- & -- & -- & \makecell{$53$\\($44$--$62$)} \\
    & \textsf{CVX-LOC}    & \makecell{$77$\\($67$--$86$)} & \makecell{$66$\\($56$--$75$)} & \makecell{$83$\\($72$--$92$)} & \makecell{$61$\\($52$--$70$)} & \makecell{$104$\\($85$--$177$)} & \makecell{$158$\\($144$--$178$)} & \makecell{$136$\\($98$--$214$)} & \makecell{$214$\\($197$--$261$)} & -- \\
    & \textsf{CVX-T} & -- & -- & -- & -- & -- & -- & -- & -- & \makecell{$3.8$\\($3.4$--$4.4$)} \\
    & \textsf{CVX-LOC-T} & \makecell{$7.8$\\($7.2$--$8.4$)} & \makecell{$8.3$\\($7.4$--$9.2$)} & \makecell{$12$\\($10$--$13$)} & \makecell{$6.9$\\($6.3$--$7.6$)} & \makecell{$18$\\($12$--$31$)} & \makecell{$100$\\($89$--$129$)} & \makecell{$20$\\($14$--$33$)} & \makecell{$153$\\($138$--$196$)} & -- \\
    & \textsf{D-LOC}    & \makecell{$121$\\($112$--$131$)} & \makecell{$48$\\($30$--$54$)} & \makecell{$257$\\($225$--$296$)} & \makecell{$40$\\($30$--$50$)} & \makecell{$125$\\($112$--$157$)} & \makecell{$393$\\($333$--$465$)} & \makecell{$94$\\($66$--$232$)} & \makecell{$226$\\($193$--$331$)} & -- \\
    & \textsf{MC}     & -- & -- & -- & -- & -- & -- & -- & -- & \makecell{$1366$\\($1320$--$1419$)} \\
    \bottomrule
    \end{tabular}}
\end{sidewaystable}

\begin{sidewaystable}
    \caption{Numerical results correspond to figure~\ref{sfig:multi_lr_2arm}.}\label{tab:multi_lr_2arm}
    \resizebox{\textwidth}{!}{%
    \begin{tabular}{@{}ccccccccccc@{}}
    \toprule
    & & Nelder-Mead & L-BFGS-B & TNC & SLSQP & Powell & Trust-Region & COBYLA & COBYQA & --\\ \midrule
    \multirow{6}{*}[-25pt]{\rotatebox{90}{\small$\expect D_{\rm kl}(\pi^{\rm gt}(t), \pi^\star(t))$}} & \textsf{CVX}    & -- & -- & -- & -- & -- & -- & -- & -- & \makecell{$0.014$\\($0.008$--$0.024$)} \\
    & \textsf{CVX-LOC}    & \makecell{$0.012$\\($0.006$--$0.021$)} & \makecell{$0.012$\\($0.006$--$0.020$)} & \makecell{$0.012$\\($0.006$--$0.020$)} & \makecell{$0.012$\\($0.006$--$0.020$)} & \makecell{$0.012$\\($0.006$--$0.020$)} & \makecell{$0.012$\\($0.006$--$0.020$)} & \makecell{$0.011$\\($0.006$--$0.020$)} & \makecell{$0.012$\\($0.006$--$0.020$)} & -- \\
    & \textsf{CVX-T} & -- & -- & -- & -- & -- & -- & -- & -- & \makecell{$0.011$\\($0.007$--$0.023$)} \\
    & \textsf{CVX-LOC-T} & \makecell{$0.012$\\($0.006$--$0.025$)} & \makecell{$0.012$\\($0.006$--$0.025$)} & \makecell{$0.012$\\($0.006$--$0.025$)} & \makecell{$0.012$\\($0.006$--$0.025$)} & \makecell{$0.012$\\($0.006$--$0.025$)} & \makecell{$0.012$\\($0.006$--$0.025$)} & \makecell{$0.013$\\($0.006$--$0.026$)} & \makecell{$0.012$\\($0.006$--$0.025$)} & -- \\
    & \textsf{D-LOC}    & \makecell{$0.014$\\($0.007$--$0.024$)} & \makecell{$0.014$\\($0.007$--$0.025$)} & \makecell{$0.014$\\($0.007$--$0.024$)} & \makecell{$0.014$\\($0.007$--$0.025$)} & \makecell{$0.014$\\($0.007$--$0.024$)} & \makecell{$0.014$\\($0.007$--$0.024$)} & \makecell{$0.014$\\($0.007$--$0.024$)} & \makecell{$0.014$\\($0.007$--$0.024$)} & -- \\
    & \textsf{MC}     & -- & -- & -- & -- & -- & -- & -- & -- & \makecell{$0.006$\\($0.003$--$0.011$)} \\ \midrule
    \multirow{5}{*}[-9pt]{\rotatebox{90}{$\norm{\alpha^{\rm gt} - \alpha^\star}_2$}} & \textsf{CVX-LOC} & \makecell{$0.30$\\($0.17$--$0.48$)} & \makecell{$0.29$\\($0.16$--$0.47$)} & \makecell{$0.29$\\($0.16$--$0.46$)} & \makecell{$0.29$\\($0.16$--$0.47$)} & \makecell{$0.29$\\($0.16$--$0.47$)} & \makecell{$0.29$\\($0.16$--$0.47$)} & \makecell{$0.29$\\($0.16$--$0.47$)} & \makecell{$0.29$\\($0.16$--$0.47$)} & -- \\
    & \textsf{CVX-LOC-T} & \makecell{$0.28$\\($0.17$--$0.43$)} & \makecell{$0.28$\\($0.16$--$0.43$)} & \makecell{$0.28$\\($0.16$--$0.43$)} & \makecell{$0.28$\\($0.16$--$0.43$)} & \makecell{$0.28$\\($0.16$--$0.43$)} & \makecell{$0.28$\\($0.17$--$0.43$)} & \makecell{$0.29$\\($0.17$--$0.45$)} & \makecell{$0.28$\\($0.17$--$0.43$)} & -- \\
    & \textsf{D-LOC}    & \makecell{$0.37$\\($0.22$--$0.61$)} & \makecell{$0.38$\\($0.22$--$0.62$)} & \makecell{$0.37$\\($0.21$--$0.61$)} & \makecell{$0.38$\\($0.22$--$0.62$)} & \makecell{$0.39$\\($0.22$--$0.64$)} & \makecell{$0.37$\\($0.21$--$0.61$)} & \makecell{$0.38$\\($0.22$--$0.63$)} & \makecell{$0.38$\\($0.22$--$0.63$)} & -- \\
    & \textsf{MC}     & -- & -- & -- & -- & -- & -- & -- & -- & \makecell{$0.22$\\($0.14$--$0.34$)} \\ \midrule
    \multirow{5}{*}[-9pt]{\rotatebox{90}{$\norm{\beta^{\rm gt} - \beta^\star}_2$}} & \textsf{CVX-LOC} & \makecell{$1.05$\\($0.60$--$1.79$)} & \makecell{$1.06$\\($0.60$--$1.80$)} & \makecell{$1.06$\\($0.60$--$1.80$)} & \makecell{$1.06$\\($0.60$--$1.80$)} & \makecell{$1.06$\\($0.60$--$1.81$)} & \makecell{$1.06$\\($0.62$--$1.81$)} & \makecell{$1.06$\\($0.60$--$1.81$)} & \makecell{$1.07$\\($0.62$--$1.84$)} & -- \\
    & \textsf{CVX-LOC-T} & \makecell{$1.12$\\($0.59$--$2.24$)} & \makecell{$1.02$\\($0.55$--$1.87$)} & \makecell{$1.07$\\($0.58$--$1.89$)} & \makecell{$1.04$\\($0.56$--$1.88$)} & \makecell{$1.02$\\($0.55$--$1.86$)} & \makecell{$1.10$\\($0.58$--$1.95$)} & \makecell{$1.14$\\($0.64$--$2.09$)} & \makecell{$1.12$\\($0.59$--$2.23$)} & -- \\
    & \textsf{D-LOC}    & \makecell{$0.98$\\($0.58$--$1.74$)} & \makecell{$1.05$\\($0.62$--$1.81$)} & \makecell{$1.04$\\($0.59$--$1.75$)} & \makecell{$1.05$\\($0.61$--$1.85$)} & \makecell{$1.07$\\($0.60$--$1.91$)} & \makecell{$1.07$\\($0.61$--$2.02$)} & \makecell{$1.08$\\($0.62$--$1.83$)} & \makecell{$1.11$\\($0.63$--$2.14$)} & -- \\
    & \textsf{MC}     & -- & -- & -- & -- & -- & -- & -- & -- & \makecell{$0.64$\\($0.40$--$1.00$)} \\ \midrule
    \multirow{6}{*}[-20pt]{\rotatebox{90}{time (ms)}} & \textsf{CVX}    & -- & -- & -- & -- & -- & -- & -- & -- & \makecell{$80$\\($64$--$97$)} \\
    & \textsf{CVX-LOC}    & \makecell{$108$\\($90$--$125$)} & \makecell{$157$\\($138$--$176$)} & \makecell{$117$\\($101$--$135$)} & \makecell{$93$\\($77$--$110$)} & \makecell{$184$\\($140$--$248$)} & \makecell{$207$\\($180$--$274$)} & \makecell{$218$\\($161$--$301$)} & \makecell{$283$\\($247$--$350$)} & -- \\
    & \textsf{CVX-T} & -- & -- & -- & -- & -- & -- & -- & -- & \makecell{$4.5$\\($3.9$--$5.3$)} \\
    & \textsf{CVX-LOC-T} & \makecell{$10$\\($9.4$--$11$)} & \makecell{$74$\\($71$--$77$)} & \makecell{$15$\\($14$--$17$)} & \makecell{$10$\\($9$--$11$)} & \makecell{$21$\\($17$--$33$)} & \makecell{$127$\\($109$--$239$)} & \makecell{$30$\\($20$--$39$)} & \makecell{$184$\\($154$--$252$)} & -- \\
    & \textsf{D-LOC}    & \makecell{$481$\\($410$--$559$)} & \makecell{$142$\\($118$--$179$)} & \makecell{$927$\\($855$--$1154$)} & \makecell{$120$\\($98$--$163$)} & \makecell{$374$\\($299$--$462$)} & \makecell{$1038$\\($875$--$1342$)} & \makecell{$521$\\($252$--$995$)} & \makecell{$1040$\\($575$--$3092$)} & -- \\
    & \textsf{MC}     & -- & -- & -- & -- & -- & -- & -- & -- & \makecell{$1811$\\($1741$--$1901$)} \\
    \bottomrule
    \end{tabular}}
\end{sidewaystable}

\begin{sidewaystable}
    \caption{Numerical results correspond to figure~\ref{sfig:subrew_2arm}.}\label{tab:subrew_2arm}
    \resizebox{\textwidth}{!}{%
    \begin{tabular}{@{}ccccccccccc@{}}
    \toprule
    & & Nelder-Mead & L-BFGS-B & TNC & SLSQP & Powell & Trust-Region & COBYLA & COBYQA & --\\ \midrule
    \multirow{6}{*}[-25pt]{\rotatebox{90}{\small$\expect D_{\rm kl}(\pi^{\rm gt}(t), \pi^\star(t))$}} & \textsf{CVX}    & -- & -- & -- & -- & -- & -- & -- & -- & \makecell{$0.022$\\($0.012$--$0.071$)} \\
    & \textsf{CVX-LOC}    & \makecell{$0.019$\\($0.010$--$0.036$)} & \makecell{$0.019$\\($0.010$--$0.036$)} & \makecell{$0.019$\\($0.010$--$0.036$)} & \makecell{$0.019$\\($0.010$--$0.036$)} & \makecell{$0.019$\\($0.010$--$0.036$)} & \makecell{$0.019$\\($0.010$--$0.036$)} & \makecell{$0.020$\\($0.010$--$0.036$)} & \makecell{$0.019$\\($0.010$--$0.036$)} & -- \\
    & \textsf{CVX-T} & -- & -- & -- & -- & -- & -- & -- & -- & \makecell{$0.018$\\($0.010$--$0.048$)} \\
    & \textsf{CVX-LOC-T} & \makecell{$0.018$\\($0.010$--$0.034$)} & \makecell{$0.018$\\($0.010$--$0.034$)} & \makecell{$0.018$\\($0.010$--$0.034$)} & \makecell{$0.018$\\($0.010$--$0.034$)} & \makecell{$0.018$\\($0.010$--$0.034$)} & \makecell{$0.018$\\($0.010$--$0.034$)} & \makecell{$0.018$\\($0.010$--$0.034$)} & \makecell{$0.018$\\($0.010$--$0.034$)} & -- \\
    & \textsf{D-LOC}    & \makecell{$0.017$\\($0.011$--$0.025$)} & \makecell{$0.017$\\($0.011$--$0.026$)} & \makecell{$0.017$\\($0.011$--$0.026$)} & \makecell{$0.017$\\($0.010$--$0.025$)} & \makecell{$0.017$\\($0.011$--$0.025$)} & \makecell{$0.017$\\($0.010$--$0.025$)} & \makecell{$0.017$\\($0.010$--$0.025$)} & \makecell{$0.017$\\($0.010$--$0.025$)} & -- \\
    & \textsf{MC}     & -- & -- & -- & -- & -- & -- & -- & -- & \makecell{$0.009$\\($0.005$--$0.014$)} \\ \midrule
    \multirow{5}{*}[-9pt]{\rotatebox{90}{$\norm{\alpha^{\rm gt} - \alpha^\star}_2$}} & \textsf{CVX-LOC} & \makecell{$0.83$\\($0.61$--$1.02$)} & \makecell{$0.73$\\($0.50$--$0.91$)} & \makecell{$0.74$\\($0.53$--$0.94$)} & \makecell{$0.74$\\($0.50$--$0.93$)} & \makecell{$0.78$\\($0.56$--$0.98$)} & \makecell{$0.78$\\($0.56$--$0.98$)} & \makecell{$0.77$\\($0.55$--$0.97$)} & \makecell{$0.78$\\($0.56$--$0.98$)} & -- \\
    & \textsf{CVX-LOC-T} & \makecell{$0.70$\\($0.50$--$0.91$)} & \makecell{$0.73$\\($0.52$--$0.94$)} & \makecell{$0.71$\\($0.50$--$0.92$)} & \makecell{$0.73$\\($0.52$--$0.94$)} & \makecell{$0.73$\\($0.51$--$0.93$)} & \makecell{$0.73$\\($0.52$--$0.93$)} & \makecell{$0.73$\\($0.52$--$0.93$)} & \makecell{$0.73$\\($0.52$--$0.93$)} & -- \\
    & \textsf{D-LOC}    & \makecell{$0.81$\\($0.60$--$1.01$)} & \makecell{$0.76$\\($0.56$--$0.97$)} & \makecell{$0.73$\\($0.53$--$0.94$)} & \makecell{$0.77$\\($0.57$--$1.00$)} & \makecell{$0.78$\\($0.57$--$0.99$)} & \makecell{$0.75$\\($0.54$--$0.96$)} & \makecell{$0.76$\\($0.56$--$0.96$)} & \makecell{$0.81$\\($0.59$--$1.01$)} & -- \\
    & \textsf{MC}     & -- & -- & -- & -- & -- & -- & -- & -- & \makecell{$0.477$\\($0.377$--$0.581$)} \\ \midrule
    \multirow{5}{*}[-9pt]{\rotatebox{90}{$\norm{\beta^{\rm gt} - \beta^\star}_2$}} & \textsf{CVX-LOC} & \makecell{$2.00$\\($1.41$--$2.80$)} & \makecell{$2.00$\\($1.40$--$2.84$)} & \makecell{$2.01$\\($1.40$--$2.83$)} & \makecell{$2.00$\\($1.41$--$2.80$)} & \makecell{$2.01$\\($1.42$--$2.84$)} & \makecell{$1.98$\\($1.35$--$2.88$)} & \makecell{$1.99$\\($1.41$--$2.77$)} & \makecell{$1.93$\\($1.32$--$2.74$)} & -- \\
    & \textsf{CVX-LOC-T} & \makecell{$2.02$\\($1.32$--$3.00$)} & \makecell{$2.11$\\($1.43$--$2.99$)} & \makecell{$2.13$\\($1.44$--$3.03$)} & \makecell{$2.06$\\($1.41$--$2.99$)} & \makecell{$2.01$\\($1.32$--$2.96$)} & \makecell{$1.98$\\($1.41$--$2.91$)} & \makecell{$1.99$\\($1.36$--$2.87$)} & \makecell{$2.13$\\($1.45$--$3.05$)} & -- \\
    & \textsf{D-LOC}    & \makecell{$2.02$\\($1.44$--$2.73$)} & \makecell{$2.01$\\($1.40$--$2.75$)} & \makecell{$1.84$\\($1.31$--$2.58$)} & \makecell{$2.05$\\($1.45$--$2.76$)} & \makecell{$2.02$\\($1.37$--$2.79$)} & \makecell{$2.05$\\($1.42$--$2.91$)} & \makecell{$1.87$\\($1.31$--$2.52$)} & \makecell{$2.14$\\($1.51$--$3.00$)} & -- \\
    & \textsf{MC}     & -- & -- & -- & -- & -- & -- & -- & -- & \makecell{$1.17$\\($0.84$--$1.59$)} \\ \midrule
    \multirow{6}{*}[-20pt]{\rotatebox{90}{time (ms)}} & \textsf{CVX}    & -- & -- & -- & -- & -- & -- & -- & -- & \makecell{$186$\\($161$--$214$)} \\
    & \textsf{CVX-LOC}    & \makecell{$215$\\($189$--$242$)} & \makecell{$301$\\($266$--$336$)} & \makecell{$227$\\($200$--$255$)} & \makecell{$205$\\($178$--$233$)} & \makecell{$350$\\($283$--$431$)} & \makecell{$325$\\($292$--$370$)} & \makecell{$357$\\($273$--$435$)} & \makecell{$415$\\($355$--$488$)} & -- \\
    & \textsf{CVX-T} & -- & -- & -- & -- & -- & -- & -- & -- & \makecell{$5.9$\\($4.9$--$7.5$)} \\
    & \textsf{CVX-LOC-T} & \makecell{$12$\\($11$--$13$)} & \makecell{$87$\\($82$--$105$)} & \makecell{$17$\\($16$--$19$)} & \makecell{$12$\\($11$--$13$)} & \makecell{$33$\\($25$--$42$)} & \makecell{$154$\\($120$--$293$)} & \makecell{$33$\\($23$--$42$)} & \makecell{$194$\\($154$--$292$)} & -- \\
    & \textsf{D-LOC}    & \makecell{$4532$\\($4060$--$4928$)} & \makecell{$1040$\\($765$--$1409$)} & \makecell{$3025$\\($2939$--$3043$)} & \makecell{$655$\\($515$--$842$)} & \makecell{$2255$\\($1966$--$2611$)} & \makecell{$5848$\\($4402$--$10247$)} & \makecell{$2741$\\($2080$--$3274$)} & \makecell{$8124$\\($3083$--$16620$)} & -- \\
    & \textsf{MC}     & -- & -- & -- & -- & -- & -- & -- & -- & \makecell{$2967$\\($2829$--$3177$)} \\
    \bottomrule
    \end{tabular}}
\end{sidewaystable}

\begin{sidewaystable}
    \caption{Numerical results correspond to figure~\ref{sfig:basic_fql_10arm}.}\label{tab:basic_fql_10arm}
    \resizebox{\textwidth}{!}{%
    \begin{tabular}{@{}ccccccccccc@{}}
    \toprule
    & & Nelder-Mead & L-BFGS-B & TNC & SLSQP & Powell & Trust-Region & COBYLA & COBYQA & --\\ \midrule
    \multirow{6}{*}[-25pt]{\rotatebox{90}{\small$\expect D_{\rm kl}(\pi^{\rm gt}(t), \pi^\star(t))$}} & \textsf{CVX}    & -- & -- & -- & -- & -- & -- & -- & -- & \makecell{$0.009$\\($0.005$--$0.026$)} \\
    & \textsf{CVX-LOC}    & \makecell{$0.005$\\($0.002$--$0.014$)} & \makecell{$0.005$\\($0.002$--$0.014$)} & \makecell{$0.005$\\($0.002$--$0.014$)} & \makecell{$0.005$\\($0.002$--$0.014$)} & \makecell{$0.005$\\($0.002$--$0.014$)} & \makecell{$0.005$\\($0.002$--$0.014$)} & \makecell{$0.005$\\($0.002$--$0.014$)} & \makecell{$0.005$\\($0.002$--$0.014$)} & -- \\
    & \textsf{CVX-T} & -- & -- & -- & -- & -- & -- & -- & -- & \makecell{$0.012$\\($0.005$--$0.037$)} \\
    & \textsf{CVX-LOC-T} & \makecell{$0.007$\\($0.003$--$0.023$)} & \makecell{$0.007$\\($0.003$--$0.023$)} & \makecell{$0.007$\\($0.003$--$0.023$)} & \makecell{$0.007$\\($0.003$--$0.023$)} & \makecell{$0.007$\\($0.003$--$0.023$)} & \makecell{$0.007$\\($0.003$--$0.023$)} & \makecell{$0.007$\\($0.003$--$0.023$)} & \makecell{$0.007$\\($0.003$--$0.023$)} & -- \\
    & \textsf{D-LOC}    & \makecell{$0.009$\\($0.003$--$0.021$)} & \makecell{$0.009$\\($0.003$--$0.021$)} & \makecell{$0.009$\\($0.003$--$0.021$)} & \makecell{$0.009$\\($0.003$--$0.021$)} & \makecell{$0.009$\\($0.003$--$0.021$)} & \makecell{$0.009$\\($0.003$--$0.021$)} & \makecell{$0.009$\\($0.003$--$0.021$)} & \makecell{$0.009$\\($0.003$--$0.021$)} & -- \\
    & \textsf{MC}     & -- & -- & -- & -- & -- & -- & -- & -- & \makecell{$0.002$\\($0.001$--$0.005$)} \\ \midrule
    \multirow{5}{*}[-9pt]{\rotatebox{90}{$|\alpha^{\rm gt} - \alpha^\star|$}} & \textsf{CVX-LOC} & \makecell{$0.07$\\($0.02$--$0.15$)} & \makecell{$0.07$\\($0.02$--$0.15$)} & \makecell{$0.07$\\($0.02$--$0.15$)} & \makecell{$0.07$\\($0.02$--$0.15$)} & \makecell{$0.07$\\($0.02$--$0.15$)} & \makecell{$0.07$\\($0.02$--$0.15$)} & \makecell{$0.07$\\($0.02$--$0.15$)} & \makecell{$0.07$\\($0.02$--$0.15$)} & -- \\
    & \textsf{CVX-LOC-T} & \makecell{$0.07$\\($0.03$--$0.15$)} & \makecell{$0.07$\\($0.03$--$0.15$)} & \makecell{$0.07$\\($0.03$--$0.15$)} & \makecell{$0.07$\\($0.03$--$0.15$)} & \makecell{$0.07$\\($0.03$--$0.15$)} & \makecell{$0.07$\\($0.03$--$0.15$)} & \makecell{$0.07$\\($0.03$--$0.15$)} & \makecell{$0.07$\\($0.03$--$0.15$)} & -- \\
    & \textsf{D-LOC}    & \makecell{$0.08$\\($0.03$--$0.20$)} & \makecell{$0.08$\\($0.03$--$0.20$)} & \makecell{$0.08$\\($0.03$--$0.20$)} & \makecell{$0.08$\\($0.03$--$0.20$)} & \makecell{$0.08$\\($0.03$--$0.20$)} & \makecell{$0.08$\\($0.03$--$0.20$)} & \makecell{$0.08$\\($0.03$--$0.20$)} & \makecell{$0.08$\\($0.03$--$0.20$)} & -- \\
    & \textsf{MC}     & -- & -- & -- & -- & -- & -- & -- & -- & \makecell{$0.05$\\($0.02$--$0.09$)} \\ \midrule
    \multirow{5}{*}[-9pt]{\rotatebox{90}{$|\beta^{\rm gt} - \beta^\star|$}} & \textsf{CVX-LOC} & \makecell{$0.72$\\($0.30$--$1.38$)} & \makecell{$0.72$\\($0.30$--$1.34$)} & \makecell{$0.71$\\($0.30$--$1.34$)} & \makecell{$0.71$\\($0.30$--$1.34$)} & \makecell{$0.71$\\($0.30$--$1.34$)} & \makecell{$0.71$\\($0.30$--$1.34$)} & \makecell{$0.69$\\($0.29$--$1.32$)} & \makecell{$0.71$\\($0.30$--$1.34$)} & -- \\
    & \textsf{CVX-LOC-T} & \makecell{$0.90$\\($0.37$--$1.81$)} & \makecell{$0.90$\\($0.37$--$1.81$)} & \makecell{$0.90$\\($0.37$--$1.81$)} & \makecell{$0.90$\\($0.37$--$1.80$)} & \makecell{$0.90$\\($0.37$--$1.80$)} & \makecell{$0.90$\\($0.38$--$1.81$)} & \makecell{$0.85$\\($0.37$--$1.72$)} & \makecell{$0.90$\\($0.37$--$1.81$)} & -- \\
    & \textsf{D-LOC}    & \makecell{$0.96$\\($0.42$--$1.67$)} & \makecell{$0.96$\\($0.42$--$1.67$)} & \makecell{$0.96$\\($0.42$--$1.67$)} & \makecell{$0.96$\\($0.42$--$1.67$)} & \makecell{$0.96$\\($0.42$--$1.67$)} & \makecell{$0.96$\\($0.42$--$1.67$)} & \makecell{$0.92$\\($0.42$--$1.65$)} & \makecell{$0.96$\\($0.42$--$1.67$)} & -- \\
    & \textsf{MC}     & -- & -- & -- & -- & -- & -- & -- & -- & \makecell{$0.47$\\($0.20$--$0.87$)} \\ \midrule
    \multirow{6}{*}[-20pt]{\rotatebox{90}{time (ms)}} & \textsf{CVX}    & -- & -- & -- & -- & -- & -- & -- & -- & \makecell{$74$\\($64$--$91$)} \\
    & \textsf{CVX-LOC}    & \makecell{$94$\\($84$--$110$)} & \makecell{$90$\\($80$--$105$)} & \makecell{$100$\\($90$--$115$)} & \makecell{$81$\\($71$--$95$)} & \makecell{$116$\\($95$--$147$)} & \makecell{$173$\\($157$--$192$)} & \makecell{$303$\\($140$--$351$)} & \makecell{$231$\\($207$--$262$)} & -- \\
    & \textsf{CVX-T} & -- & -- & -- & -- & -- & -- & -- & -- & \makecell{$23$\\($20$--$26$)} \\
    & \textsf{CVX-LOC-T} & \makecell{$26$\\($24$--$29$)} & \makecell{$28$\\($25$--$30$)} & \makecell{$29$\\($27$--$32$)} & \makecell{$26$\\($24$--$29$)} & \makecell{$33$\\($29$--$38$)} & \makecell{$113$\\($101$--$123$)} & \makecell{$60$\\($37$--$68$)} & \makecell{$168$\\($134$--$185$)} & -- \\
    & \textsf{D-LOC}    & \makecell{$111$\\($86$--$124$)} & \makecell{$61$\\($53$--$73$)} & \makecell{$248$\\($181$--$289$)} & \makecell{$45$\\($39$--$54$)} & \makecell{$140$\\($111$--$235$)} & \makecell{$438$\\($321$--$509$)} & \makecell{$356$\\($136$--$921$)} & \makecell{$240$\\($209$--$288$)} & -- \\
    & \textsf{MC}     & -- & -- & -- & -- & -- & -- & -- & -- & \makecell{$2259$\\($2187$--$2334$)} \\
    \bottomrule
    \end{tabular}}
\end{sidewaystable}

\begin{sidewaystable}
    \caption{Numerical results correspond to figure~\ref{sfig:multi_lr_10arm}.}\label{tab:multi_lr_10arm}
    \resizebox{\textwidth}{!}{%
    \begin{tabular}{@{}ccccccccccc@{}}
    \toprule
    & & Nelder-Mead & L-BFGS-B & TNC & SLSQP & Powell & Trust-Region & COBYLA & COBYQA & --\\ \midrule
    \multirow{6}{*}[-25pt]{\rotatebox{90}{\small$\expect D_{\rm kl}(\pi^{\rm gt}(t), \pi^\star(t))$}} & \textsf{CVX}    & -- & -- & -- & -- & -- & -- & -- & -- & \makecell{$0.09$\\($0.04$--$0.18$)} \\
    & \textsf{CVX-LOC}    & \makecell{$0.034$\\($0.017$--$0.059$)} & \makecell{$0.034$\\($0.017$--$0.059$)} & \makecell{$0.034$\\($0.017$--$0.059$)} & \makecell{$0.034$\\($0.017$--$0.059$)} & \makecell{$0.034$\\($0.017$--$0.059$)} & \makecell{$0.034$\\($0.017$--$0.059$)} & \makecell{$0.034$\\($0.017$--$0.059$)} & \makecell{$0.034$\\($0.017$--$0.059$)} & -- \\
    & \textsf{CVX-T} & -- & -- & -- & -- & -- & -- & -- & -- & \makecell{$0.08$\\($0.04$--$0.17$)} \\
    & \textsf{CVX-LOC-T} & \makecell{$0.037$\\($0.017$--$0.068$)} & \makecell{$0.036$\\($0.017$--$0.068$)} & \makecell{$0.036$\\($0.017$--$0.068$)} & \makecell{$0.036$\\($0.017$--$0.068$)} & \makecell{$0.037$\\($0.017$--$0.068$)} & \makecell{$0.036$\\($0.017$--$0.068$)} & \makecell{$0.037$\\($0.017$--$0.068$)} & \makecell{$0.036$\\($0.017$--$0.068$)} & -- \\
    & \textsf{D-LOC}    & \makecell{$0.033$\\($0.016$--$0.054$)} & \makecell{$0.034$\\($0.016$--$0.054$)} & \makecell{$0.033$\\($0.016$--$0.055$)} & \makecell{$0.034$\\($0.016$--$0.054$)} & \makecell{$0.033$\\($0.016$--$0.053$)} & \makecell{$0.034$\\($0.016$--$0.054$)} & \makecell{$0.034$\\($0.016$--$0.056$)} & \makecell{$0.034$\\($0.016$--$0.056$)} & -- \\
    & \textsf{MC}     & -- & -- & -- & -- & -- & -- & -- & -- & \makecell{$0.010$\\($0.006$--$0.018$)} \\ \midrule
    \multirow{5}{*}[-9pt]{\rotatebox{90}{$\norm{\alpha^{\rm gt} - \alpha^\star}_2$}} & \textsf{CVX-LOC} & \makecell{$1.16$\\($0.91$--$1.45$)} & \makecell{$1.16$\\($0.91$--$1.45$)} & \makecell{$1.16$\\($0.91$--$1.45$)} & \makecell{$1.16$\\($0.91$--$1.45$)} & \makecell{$1.16$\\($0.91$--$1.45$)} & \makecell{$1.16$\\($0.91$--$1.45$)} & \makecell{$1.16$\\($0.91$--$1.45$)} & \makecell{$1.16$\\($0.91$--$1.45$)} & -- \\
    & \textsf{CVX-LOC-T} & \makecell{$1.08$\\($0.85$--$1.37$)} & \makecell{$1.08$\\($0.85$--$1.37$)} & \makecell{$1.08$\\($0.85$--$1.37$)} & \makecell{$1.08$\\($0.85$--$1.37$)} & \makecell{$1.08$\\($0.85$--$1.38$)} & \makecell{$1.08$\\($0.85$--$1.37$)} & \makecell{$1.08$\\($0.85$--$1.37$)} & \makecell{$1.08$\\($0.85$--$1.37$)} & -- \\
    & \textsf{D-LOC}    & \makecell{$1.34$\\($1.13$--$1.53$)} & \makecell{$1.19$\\($1.02$--$1.37$)} & \makecell{$1.21$\\($1.02$--$1.37$)} & \makecell{$1.18$\\($1.01$--$1.37$)} & \makecell{$1.46$\\($1.24$--$1.67$)} & \makecell{$1.02$\\($0.86$--$1.18$)} & \makecell{$1.20$\\($1.03$--$1.39$)} & \makecell{$1.31$\\($1.11$--$1.51$)} & -- \\
    & \textsf{MC}     & -- & -- & -- & -- & -- & -- & -- & -- & \makecell{$0.71$\\($0.58$--$0.83$)} \\ \midrule
    \multirow{5}{*}[-9pt]{\rotatebox{90}{$\norm{\beta^{\rm gt} - \beta^\star}_2$}} & \textsf{CVX-LOC} & \makecell{$8.17$\\($7.18$--$9.00$)} & \makecell{$8.18$\\($7.18$--$9.02$)} & \makecell{$8.15$\\($7.13$--$8.98$)} & \makecell{$8.17$\\($7.18$--$9.02$)} & \makecell{$8.17$\\($7.18$--$9.02$)} & \makecell{$8.19$\\($7.23$--$9.01$)} & \makecell{$8.00$\\($7.08$--$8.92$)} & \makecell{$8.01$\\($7.02$--$8.93$)} & -- \\
    & \textsf{CVX-LOC-T} & \makecell{$7.92$\\($7.00$--$8.88$)} & \makecell{$8.08$\\($7.14$--$8.98$)} & \makecell{$7.96$\\($7.00$--$8.85$)} & \makecell{$8.08$\\($7.13$--$8.96$)} & \makecell{$8.02$\\($7.10$--$8.94$)} & \makecell{$8.18$\\($7.24$--$8.99$)} & \makecell{$8.01$\\($7.01$--$8.87$)} & \makecell{$8.25$\\($7.27$--$9.08$)} & -- \\
    & \textsf{D-LOC}    & \makecell{$7.66$\\($6.68$--$8.52$)} & \makecell{$6.65$\\($5.69$--$7.55$)} & \makecell{$6.42$\\($5.56$--$7.38$)} & \makecell{$6.64$\\($5.78$--$7.50$)} & \makecell{$8.22$\\($7.18$--$9.19$)} & \makecell{$5.94$\\($5.08$--$6.84$)} & \makecell{$6.17$\\($5.40$--$7.04$)} & \makecell{$6.77$\\($5.86$--$7.68$)} & -- \\
    & \textsf{MC}     & -- & -- & -- & -- & -- & -- & -- & -- & \makecell{$4.21$\\($3.69$--$4.67$)} \\ \midrule
    \multirow{6}{*}[-20pt]{\rotatebox{90}{time (ms)}} & \textsf{CVX}    & -- & -- & -- & -- & -- & -- & -- & -- & \makecell{$171$\\($120$--$408$)} \\
    & \textsf{CVX-LOC}    & \makecell{$201$\\($142$--$450$)} & \makecell{$343$\\($224$--$634$)} & \makecell{$213$\\($152$--$459$)} & \makecell{$190$\\($134$--$443$)} & \makecell{$262$\\($189$--$514$)} & \makecell{$1209$\\($443$--$1412$)} & \makecell{$442$\\($348$--$716$)} & \makecell{$444$\\($354$--$738$)} & -- \\
    & \textsf{CVX-T} & -- & -- & -- & -- & -- & -- & -- & -- & \makecell{$29$\\($26$--$32$)} \\
    & \textsf{CVX-LOC-T} & \makecell{$35$\\($32$--$39$)} & \makecell{$237$\\($147$--$329$)} & \makecell{$42$\\($38$--$45$)} & \makecell{$34$\\($32$--$38$)} & \makecell{$46$\\($42$--$53$)} & \makecell{$205$\\($181$--$656$)} & \makecell{$69$\\($57$--$76$)} & \makecell{$222$\\($187$--$257$)} & -- \\
    & \textsf{D-LOC}    & \makecell{$7595$\\($7194$--$7646$)} & \makecell{$1425$\\($762$--$2284$)} & \makecell{$7626$\\($5206$--$7871$)} & \makecell{$814$\\($493$--$1104$)} & \makecell{$2394$\\($2101$--$2794$)} & \makecell{$8810$\\($5248$--$15874$)} & \makecell{$1927$\\($1859$--$1956$)} & \makecell{$18286$\\($5331$--$41791$)} & -- \\
    & \textsf{MC}     & -- & -- & -- & -- & -- & -- & -- & -- & \makecell{$3828$\\($3573$--$4111$)} \\
    \bottomrule
    \end{tabular}}
\end{sidewaystable}

\begin{sidewaystable}
    \caption{Numerical results correspond to figure~\ref{sfig:subrew_10arm}.}\label{tab:subrew_10arm}
    \resizebox{\textwidth}{!}{%
    \begin{threeparttable}
    \begin{tabular}{@{}ccccccccccc@{}}
    \toprule
    & & Nelder-Mead & L-BFGS-B & TNC & SLSQP & Powell & Trust-Region & COBYLA & COBYQA\tnote{*} & --\\ \midrule
    \multirow{6}{*}[-25pt]{\rotatebox{90}{\small$\expect D_{\rm kl}(\pi^{\rm gt}(t), \pi^\star(t))$}} & \textsf{CVX}    & -- & -- & -- & -- & -- & -- & -- & -- & \makecell{$0.078$\\($0.032$--$0.174$)} \\
    & \textsf{CVX-LOC}    & \makecell{$0.030$\\($0.010$--$0.075$)} & \makecell{$0.031$\\($0.011$--$0.075$)} & \makecell{$0.031$\\($0.011$--$0.075$)} & \makecell{$0.031$\\($0.011$--$0.075$)} & \makecell{$0.031$\\($0.011$--$0.075$)} & \makecell{$0.031$\\($0.011$--$0.075$)} & \makecell{$0.031$\\($0.010$--$0.075$)} & \makecell{$0.031$\\($0.011$--$0.075$)} & -- \\
    & \textsf{CVX-T} & -- & -- & -- & -- & -- & -- & -- & -- & \makecell{$0.075$\\($0.028$--$0.167$)} \\
    & \textsf{CVX-LOC-T} & \makecell{$0.026$\\($0.010$--$0.056$)} & \makecell{$0.026$\\($0.010$--$0.056$)} & \makecell{$0.026$\\($0.010$--$0.056$)} & \makecell{$0.026$\\($0.010$--$0.056$)} & \makecell{$0.026$\\($0.010$--$0.056$)} & \makecell{$0.026$\\($0.010$--$0.056$)} & \makecell{$0.026$\\($0.010$--$0.057$)} & \makecell{$0.026$\\($0.010$--$0.056$)} & -- \\
    & \textsf{D-LOC}    & \makecell{$0.023$\\($0.011$--$0.039$)} & \makecell{$0.024$\\($0.011$--$0.041$)} & \makecell{$0.023$\\($0.011$--$0.039$)} & \makecell{$0.024$\\($0.011$--$0.041$)} & \makecell{$0.023$\\($0.011$--$0.039$)} & \makecell{$0.024$\\($0.011$--$0.040$)} & \makecell{$0.022$\\($0.010$--$0.037$)} & -- & -- \\
    & \textsf{MC}     & -- & -- & -- & -- & -- & -- & -- & -- & \makecell{$0.009$\\($0.003$--$0.017$)} \\ \midrule
    \multirow{5}{*}[-9pt]{\rotatebox{90}{$\norm{\alpha^{\rm gt} - \alpha^\star}_2$}} & \textsf{CVX-LOC} & \makecell{$2.31$\\($2.05$--$2.52$)} & \makecell{$2.26$\\($2.02$--$2.48$)} & \makecell{$2.24$\\($2.00$--$2.47$)} & \makecell{$2.26$\\($2.02$--$2.48$)} & \makecell{$2.27$\\($2.03$--$2.48$)} & \makecell{$2.26$\\($2.03$--$2.48$)} & \makecell{$2.26$\\($2.02$--$2.48$)} & \makecell{$2.26$\\($2.02$--$2.48$)} & -- \\
    & \textsf{CVX-LOC-T} & \makecell{$2.19$\\($1.90$--$2.44$)} & \makecell{$2.21$\\($1.94$--$2.46$)} & \makecell{$2.15$\\($1.86$--$2.42$)} & \makecell{$2.21$\\($1.94$--$2.46$)} & \makecell{$2.21$\\($1.94$--$2.46$)} & \makecell{$2.21$\\($1.94$--$2.46$)} & \makecell{$2.18$\\($1.92$--$2.45$)} & \makecell{$2.21$\\($1.94$--$2.46$)} & -- \\
    & \textsf{D-LOC}    & \makecell{$2.17$\\($1.93$--$2.37$)} & \makecell{$1.89$\\($1.71$--$2.06$)} & \makecell{$1.85$\\($1.68$--$2.04$)} & \makecell{$1.88$\\($1.70$--$2.06$)} & \makecell{$2.34$\\($2.16$--$2.54$)} & \makecell{$1.58$\\($1.40$--$1.77$)} & \makecell{$1.90$\\($1.70$--$2.07$)} & -- & -- \\
    & \textsf{MC}     & -- & -- & -- & -- & -- & -- & -- & -- & \makecell{$1.22$\\($1.12$--$1.32$)} \\ \midrule
    \multirow{5}{*}[-9pt]{\rotatebox{90}{$\norm{\beta^{\rm gt} - \beta^\star}_2$}} & \textsf{CVX-LOC} & \makecell{$12.2$\\($11.2$--$13.1$)} & \makecell{$12.1$\\($11.1$--$13.0$)} & \makecell{$11.9$\\($11.0$--$12.9$)} & \makecell{$12.0$\\($11.0$--$12.9$)} & \makecell{$12.2$\\($11.3$--$13.1$)} & \makecell{$11.9$\\($11.0$--$12.9$)} & \makecell{$12.1$\\($11.2$--$13.0$)} & \makecell{$11.8$\\($10.8$--$12.8$)} & -- \\
    & \textsf{CVX-LOC-T} & \makecell{$12.2$\\($11.1$--$13.2$)} & \makecell{$11.7$\\($10.7$--$12.7$)} & \makecell{$12.0$\\($11.0$--$13.0$)} & \makecell{$11.9$\\($11.0$--$13.0$)} & \makecell{$12.1$\\($11.1$--$13.2$)} & \makecell{$11.7$\\($10.8$--$12.8$)} & \makecell{$12.0$\\($11.0$--$13.0$)} & \makecell{$11.8$\\($10.8$--$12.8$)} & -- \\
    & \textsf{D-LOC}    & \makecell{$10.9$\\($9.8$--$12.0$)} & \makecell{$9.2$\\($8.3$--$10.2$)} & \makecell{$9.2$\\($8.4$--$10.1$)} & \makecell{$9.3$\\($8.4$--$10.1$)} & \makecell{$12.0$\\($11.0$--$13.0$)} & \makecell{$7.8$\\($7.1$--$8.7$)} & \makecell{$9.0$\\($8.2$--$9.9$)} & -- & -- \\
    & \textsf{MC}     & -- & -- & -- & -- & -- & -- & -- & -- & \makecell{$6.25$\\($5.78$--$6.72$)} \\ \midrule
    \multirow{6}{*}[-20pt]{\rotatebox{90}{time (ms)}} & \textsf{CVX}    & -- & -- & -- & -- & -- & -- & -- & -- & \makecell{$283$\\($241$--$374$)} \\
    & \textsf{CVX-LOC}    & \makecell{$312$\\($271$--$410$)} & \makecell{$456$\\($353$--$646$)} & \makecell{$325$\\($284$--$421$)} & \makecell{$301$\\($259$--$397$)} & \makecell{$643$\\($542$--$747$)} & \makecell{$572$\\($473$--$755$)} & \makecell{$469$\\($341$--$614$)} & \makecell{$549$\\($470$--$728$)} & -- \\
    & \textsf{CVX-T} & -- & -- & -- & -- & -- & -- & -- & -- & \makecell{$33$\\($29$--$38$)} \\
    & \textsf{CVX-LOC-T} & \makecell{$41$\\($37$--$46$)} & \makecell{$198$\\($142$--$348$)} & \makecell{$51$\\($47$--$56$)} & \makecell{$41$\\($37$--$46$)} & \makecell{$62$\\($54$--$71$)} & \makecell{$260$\\($219$--$610$)} & \makecell{$69$\\($49$--$78$)} & \makecell{$258$\\($229$--$313$)} & -- \\
    & \textsf{D-LOC}    & \makecell{$28473$\\($28010$--$28687$)} & \makecell{$5942$\\($2795$--$11515$)} & \makecell{$49086$\\($20681$--$56594$)} & \makecell{$3669$\\($2488$--$5101$)} & \makecell{$12386$\\($8115$--$15662$)} & \makecell{$53444$\\($17392$--$108439$)} & \makecell{$3854$\\($3720$--$3885$)} & -- & -- \\
    & \textsf{MC}     & -- & -- & -- & -- & -- & -- & -- & -- & \makecell{$13044$\\($12232$--$13640$)} \\
    \bottomrule
    \end{tabular}
    \begin{tablenotes}
    \item[*] The COBYQA solver could not return a solution in acceptable time when applied in the \textsf{D-LOC} method, and hence the corresponding numerical results are omitted.
    \end{tablenotes}
    \end{threeparttable}}
\end{sidewaystable}

\newpage
\bibliography{refs}

\end{document}